\documentclass[twocolumn,twocolappendix]{aastex6}
\usepackage{amsmath}
\usepackage{units}

\newcommand{\E}[1]{\times10^{#1}}
\newcommand{\msol}{ \, M_\sun }
\newcommand{\bi}{\begin{itemize}}
\newcommand{\ei}{\end{itemize}}
\newcommand{\MESA}{\texttt{MESA}}
\newcommand{\FLASH}{\texttt{FLASH}}

\shorttitle{Sub-$M_{\rm Ch}$ WD detonations}
\shortauthors{SHEN ET AL.}
\slugcomment{Accepted}

\begin{document}

\title{Sub-Chandrasekhar-mass white dwarf detonations revisited}

\author{Ken J.\ Shen\altaffilmark{1,2}, Daniel Kasen\altaffilmark{1,2,3}, Broxton J.\ Miles\altaffilmark{4}, and Dean M.\ Townsley\altaffilmark{4}}
\altaffiltext{1}{Department of Astronomy and Theoretical Astrophysics Center, University of California, Berkeley, CA, USA; kenshen@astro.berkeley.edu}
\altaffiltext{2}{Lawrence Berkeley National Laboratory, Berkeley, CA, USA}
\altaffiltext{3}{Department of Physics, University of California, Berkeley, CA, USA}
\altaffiltext{4}{Department of Physics and Astronomy, University of Alabama, Tuscaloosa, AL, USA}

\begin{abstract}

The detonation of a sub-Chandrasekhar-mass white dwarf (WD) has emerged as one of the most promising Type Ia supernova (SN Ia) progenitor scenarios.  Recent studies have suggested that the rapid transfer of a very small amount of helium from one WD to another is sufficient to ignite a helium shell detonation that subsequently triggers a carbon core detonation, yielding a ``dynamically driven double degenerate double detonation'' SN Ia.  Because the helium shell that surrounds the core explosion is so minimal, this scenario approaches the limiting case of a bare C/O WD detonation.  Motivated by discrepancies in previous literature and by a recent need for detailed nucleosynthetic data, we revisit simulations of naked C/O WD detonations in this paper.  We disagree to some extent with the nucleosynthetic results of previous work on sub-Chandrasekhar-mass bare C/O WD detonations; e.g., we find that a median-brightness SN Ia is produced by the detonation of a $1.0 \msol$ WD instead of a more massive and rarer $1.1 \msol$ WD.  The neutron-rich nucleosynthesis in our simulations agrees broadly with some observational constraints, although tensions remain with others.  There are also discrepancies related to the velocities of the outer ejecta and light curve shapes, but overall our synthetic light curves and spectra are roughly consistent with observations.  We are hopeful that future multi-dimensional simulations will resolve these  issues and further bolster the dynamically driven double degenerate double detonation scenario's potential to explain most SNe Ia.

\end{abstract}



\section{Introduction}
\label{sec:intro}

The nature of Type Ia supernova (SN Ia) progenitors remains one of the  enduring mysteries of astrophysics (for recent reviews, see \citealt{hill13a} and \citealt{maoz14a}).  For decades, many researchers favored a scenario involving a C/O white dwarf (WD) whose mass approaches the Chandrasekhar limit via stable hydrogen-rich accretion from a non-degenerate companion \citep{wi73,nomo82a} or in an unstable merger with another C/O WD \citep{it84,webb84}.  Carbon fusion at the center of the WD would then lead to a phase of convective simmering, followed by the birth of a deflagration, a transition to a detonation, and subsequently, a SN Ia explosion (e.g., \citealt{kow97,plew04a,seit13b}).

However, growing constraints from recent theoretical and observational work have increased persisting doubts that the Chandrasekhar-mass ($M_{\rm Ch}$) scenario is responsible for the bulk of SNe Ia (e.g., \citealt{leon07,sb07,kerz09a,rbf09,kase10,bloo12,sp12,wood13a,scal14b,joha16a,dhaw17a}).  Increased attention is being paid to alternative solutions, chief among them the double detonation scenario.  In its earliest incarnations \citep{wtw86,nomo82b,livn90}, this scenario invoked accretion from a non-degenerate helium-burning star onto a C/O WD, which leads to a $\sim 0.1 \msol$ helium shell that ignites, begins to convect, and then detonates.  The helium shell detonation then triggers a detonation in the sub-$M_{\rm Ch}$ C/O core via a direct edge-lit detonation or via shock convergence near the center.  However, the helium detonation in the massive shells of these early models produced $^{56}$Ni and other iron-group elements in the outer regions of the SN ejecta, which presented problems when compared to observations \citep{hk96,nuge97}.

In recent years, the realization that stable accretion from helium WD donors yields much smaller helium shells at ignition due to the higher accretion rates \citep{bild07,sb09b}, coupled with the problems besetting $M_{\rm Ch}$ scenarios, motivated a resurgence of double detonation studies focused on the explosion of sub-$M_{\rm Ch}$ WDs \citep{fhr07,fink10,krom10,wk11,shen14a}.  In parallel work, studies of unstable double WD mergers uncovered the possibility that helium could detonate as it was transferred during the dynamical phase of the merger \citep{guil10,rask12,pakm13a,moll14a,tani15a}.  This scenario was made even more attractive due to work that showed that including a large nuclear reaction network and realistic C/O pollution in the helium shell drastically reduces the minimum hotspot size and shell mass for helium detonation initiation and propagation \citep{shen14b}.

Observational studies have begun to narrow the highly uncertain double WD interaction rate, finding rough agreement with binary population synthesis calculations (e.g., \citealt{ruit11,toon17a}).  A recent observational estimate \citep{maoz17a} finds that the rate of double WDs coming into mass transfer contact is $\sim 10$ times the SN Ia rate.  Not all of these binaries necessarily lead to double WD mergers, but \cite{shen15a} introduced the possibility that all double WD systems  do indeed merge unstably due to dynamical friction during the initial phases of stable hydrogen- and helium-rich mass transfer.  Thus,  double WD binaries have the potential to explain all SNe Ia if just $\sim 10\%$ of double WD mergers lead to SNe Ia via double detonations (or via direct carbon ignition; \citealt{pakm10,pakm11,pakm12b,kash15a}).  Furthermore, prompt detonations in merging double WD binaries also have the capacity to explain the evolution of the SN Ia luminosity function \citep{shen17c}.

\cite{sim10} provided a baseline for radiative transfer simulations of double detonation SNe Ia by calculating the explosion and appearance of a bare C/O WD core with no overlying helium shell.  They found reasonable agreement with observations of SNe Ia, both in terms of light curves and spectra.  However, recent work by \cite{moll14a} included a set of hydrodynamical explosions of bare C/O WDs that disagreed with the nucleosynthetic results of \cite{sim10}.  Moreover, recent observational results concerning neutron-rich isotopes in SNe Ia (e.g., \citealt{seit13a,yama15b,dimi17a}) have been claimed as evidence against sub-$M_{\rm Ch}$ explosions, but comprehensive in-depth studies of nucleosynthetic abundances in sub-$M_{\rm Ch}$ detonations do not yet exist in the literature for comparison to these observations.

Motivated by the disagreement in previous work and by the need for detailed nucleosynthetic data, we revisit simplified simulations of spherically symmetric bare C/O WD detonations in this paper.  While recent studies have performed hydrodynamical and radiative transfer simulations with multi-dimensional helium shell ignitions (e.g., \citealt{fhr07,fink10,krom10,sim12,moll13a}), their use of relatively massive helium shells yielded significant amounts of iron-group elements in the helium detonation ashes, which continues to be a vexing issue for obtaining spectroscopically normal SNe Ia from these models.  The much smaller helium shells at ignition found by \cite{shen14b} suggest that the study of one-dimensional baseline bare WD core detonations with no helium shell is still informative. Future work will continue the development of double detonation models by including these very low mass helium shells in multi-dimensional simulations.

We begin in \S \ref{sec:setup} by describing our method for artificially broadening detonations in WDs into structures that are spatially resolved on our numerical grid.  In \S \ref{sec:results}, we detail our nucleosynthetic results for a suite of 80 post-processed simulations, focusing on bulk yields in \S \ref{sec:bulk} and on neutron-rich nucleosynthesis in \S \ref{sec:nrich}.  We perform radiative transfer simulations and demonstrate qualitative agreement with light curves (\S \ref{sec:lcs}) and spectra (\S \ref{sec:spectra}) of observed SNe Ia, and we conclude with avenues for future research in \S \ref{sec:conc}.


\section{Simulation Details}
\label{sec:setup}

In this section, we describe our procedure for setting up, running, and post-processing our reactive hydrodynamic simulations.  We begin by calculating the initial conditions for our white dwarfs (WDs) with the stellar evolution code \MESA\footnote{\texttt{http://mesa.sourceforge.net}, version 8845; default options used unless otherwise noted.} \citep{paxt11,paxt13,paxt15a}.  We construct WDs with masses of $0.8$, $0.85$, $0.9$, $1.0$, and $1.1 \msol$ and uniform compositions of $50/50$ or $30/70$ C/O by mass.  The WDs are initially hot and are allowed to cool until their central temperatures reach $\unit[3\E{7}]{K}$.

The density profiles of these 10 models are then used as initial conditions for our \FLASH\footnote{\texttt{http://flash.uchicago.edu}, version 4.2.2; default options used unless otherwise noted.} simulations \citep{fryx00,dube09a}.  \FLASH\ and \MESA\ use the same equation of state for most of the relevant parameter space \citep{ts00b}, but there is still a small deviation from hydrostatic balance in the outer regions of the WD after mapping to \FLASH.  However, any spurious velocities are erased after the detonation passes. Each one-dimensional spherically symmetric simulation has a domain size of $\unit[10^{11}]{cm}$ and 19 levels of adaptive mesh refinement for a minimum cell size of  $\unit[4.8\E{4}]{cm}$ within the WD.  The criteria for refinement are based on the gradients of pressure, density, and temperature using \FLASH's default thresholds.  At a radius initially just outside the WD's surface, the minimum allowed cell size increases by a factor of two and continues to increase linearly with radius beyond this location.  This limits the amount of computational time spent following the shock that propagates outwards into the ambient medium after the detonation passes through the WD.  Additionally, the maximum level of refinement in the innermost $\unit[10^7]{cm}$ is reduced by four levels so that inwardly propagating shocks do not limit the global timestep as they converge towards the center and increase their velocity.

The C/O ratio of the WD in \FLASH\ is set to match the \MESA\ model from which it came.  Furthermore, we include four different metallicities for our initial models: $0$, $0.5$, $1$, and $2 \, Z_\odot$, which we approximate by including $^{22}$Ne, the stopping point for CNO isotopes following helium-burning, and $^{56}$Fe at mass fractions of $X_{\rm 22Ne}=0$, $0.005$, $0.01$, and $0.02$ with $X_{\rm 56Fe}=0.1 \, X_{\rm 22Ne}$.   The ambient medium surrounding the WD is initialized with a density and temperature of $\unit[10^{-3}]{g \, cm^{-3}}$ and $\unit[10^6]{K}$.  We enable monopole gravity and nuclear burning.  Burning in physical detonations occurs behind the nearly infinitesimally thin shock front, so reactions in \FLASH\ are disabled within shocks by default to avoid unphysical detonation structures.  \cite{fma89a} showed that an Eulerian piecewise parabolic method hydrodynamics code with reactions disabled within shocks produces the correct detonation speeds even for unresolved burning, as well as avoiding a potential artificial deflagration caused by numerical mixing at the shock front.  See Appendix A of \cite{town16a} for a more detailed discussion.

We have extended \FLASH's nuclear burning capabilities by incorporating an interface to \MESA's nuclear burning module, which enables the ability to construct an arbitrary nuclear reaction network.  For our hydrodynamic simulations, we use a 41-isotope network comprised of neutrons, $^1$H, $^4$He, $^{11}$B, $^{12}$C, $^{13-14}$N, $^{16-17}$O, $^{20,22}$Ne, $^{23}$Na, $^{24-26}$Mg, $^{27}$Al, $^{28-30}$Si, $^{30-31}$P, $^{31-32}$S, $^{35}$Cl, $^{36-39}$Ar, $^{39}$K, $^{40}$Ca, $^{43}$Sc, $^{44}$Ti, $^{47}$V, $^{48}$Cr, $^{51}$Mn, $^{52,56}$Fe, $^{55}$Co, and $^{56,58-59}$Ni, with 190 interlinking reactions from JINA's \texttt{REACLIB} \citep{cybu10a}.  For the relevant detonation conditions in C/O-rich material, this network yields errors of at most a few percent in the energy release.  Note that the above network is only tailored to track accurate energy release but not accurate isotopic abundances.  In order to more precisely calculate abundances, we include tracer particles for post-processing, which track the radius, velocity, density, and temperature and are evenly spaced every $\unit[5\E{6}]{cm}$ throughout the WD.

The detonation is ignited at the center of the WD by initializing a hotspot of radius $\unit[4\E{7}]{cm}$ that has a linear temperature gradient with a central temperature of $\unit[2\E{9}]{K}$ and an outer temperature of $\unit[1.2\E{9}]{K}$.  The temperature just outside the hotspot and throughout the rest of the WD is set to a constant $\unit[3\E{7}]{K}$; note that the value of the initial WD temperature is unimportant because post-shock temperatures are $\sim 100$ times higher.  The $\unit[4\E{7}]{cm}$  hotspot is much larger than the minimum detonatable regions found by previous work \citep{al94b,nw97,rwh07a,seit09} but is necessary due to the burning limiter we describe below.  We have confirmed that our results are insensitive to the size of the hotspot, which is reasonable because inaccurate nucleosynthesis due to the temperature perturbation will be confined to the hotspot, which corresponds to a central mass of just $0.0014$, $0.0018$, $0.0025$, $0.0045$, and $0.0088 \msol$ for our $0.8$, $0.85$, $0.9$, $1.0$, and $1.1 \msol$ $50/50$ C/O WDs.

One goal of our work is to ensure that we are capturing the relevant  physics by spatially resolving the reaction front structure in our simulations.  However, C/O detonations have lengthscales $\unit[10-10^4]{cm}$ at our densities of interest and have thus been previously followed with a level set or progress variable method (e.g., \citealt{cald07a,sim10,seit13b,town16a}).  We overcome this obstacle by artificially broadening the detonation, similar in spirit to previous studies that thicken deflagration fronts \citep{khok95a,cald07a,town16a}, and subsequently testing our resolved simulations for convergence.  We broaden the detonation by introducing a limit to the relative amount the temperature can change within each cell in one timestep due to nuclear burning, $ | \Delta \ln T |_{\rm max}$, similar to the method employed by \cite{kush13a}.  For our primary simulations we choose $ | \Delta \ln T |_{\rm max} = 0.04$, motivated by the convergence studies detailed in \S \ref{sec:conv}.

Reactions and hydrodynamics in \FLASH\ are computed in an operator-split fashion.  Between each computation of the hydrodynamic evolution during a time $\Delta t_{\rm hydro}$, the timestep determined by the Courant condition \citep{fryx00}, a temporally resolved integration of the reactions is performed in each cell with an initial integration time of $\Delta t_{\rm react}=\Delta t_{\rm hydro}$.  From the entropy equation, the relative change in temperature is $ \Delta \ln T \sim \bar{  \epsilon }  \Delta t_{\rm react} / c_V T $, where $ \bar{ \epsilon }$ is the average energy generation rate over the timestep $\Delta t_{\rm react}$, and $c_V$ is the specific heat at constant volume.  If $ | \Delta \ln T | >  | \Delta \ln T |_{\rm max}$ in a cell, the burning integration time, $ \Delta t_{\rm react}$, is reduced to the appropriate value via a Newton-Raphson iteration while leaving $ \Delta t_{\rm hydro}$ unchanged.  The burning integration is re-run for each iteration in order to yield consistent energetics and abundance changes.   This limiting procedure can also be thought of as integrating the reactions for the full hydrodynamic timestep, $\Delta t_{\rm hydro}$, but with all the reaction rates reduced by the multiplicative factor $\Delta t_{\rm react} / \Delta t_{\rm hydro}$.

Simulations are evolved for $\unit[10]{s}$, after which the tracer particles' density and temperature histories are post-processed with \MESA's one zone burner.  We employ a $205$-isotope network that includes neutrons, $^{1-2}$H, $^{3-4}$He, $^{6-7}$Li, $^{7,9-10}$Be, $^{8,10-11}$B, $^{12-13}$C, $^{13-16}$N, $^{15-19}$O, $^{17-20}$F, $^{19-23}$Ne, $^{21-24}$Na, $^{23-27}$Mg, $^{25-28}$Al, $^{27-33}$Si, $^{30-34}$P, $^{31-37}$S, $^{35-38}$Cl, $^{35-41}$Ar, $^{39-44}$K, $^{39-49}$Ca, $^{43-51}$Sc, $^{43-54}$Ti, $^{47-56}$V, $^{47-58}$Cr, $^{51-56}$Mn, $^{51-62}$Fe, $^{54-62}$Co, $^{54-62}$Ni, $^{58-66}$Cu, $^{59-66}$Zn, $^{59-66}$Ga, and $^{59-66}$Ge and interlinking reactions from JINA's \texttt{REACLIB} \citep{cybu10a}. We post-process each of our 40 hydrodynamic simulations with two different normalizations of the $^{12}$C$+^{16}$O reaction rate ($1$ and $0.1$ times the default rate; see \S \ref{sec:SNR_CrFeCaS} for the motivation behind this variation) for a total of 80 post-processed results.


\subsection{Convergence studies}
\label{sec:conv}

In this section, we demonstrate the convergence of our results as we increase the resolution in our simulations for a set of $1.0 \msol$ $50/50$ C/O solar metallicity WD detonations.  Note that since the physical burning scales are not resolved by many orders of magnitude, convergence does not imply correctness, only that our thickening scheme is numerically consistent over the range of grid scales used here. Verification of yields against resolved calculations will be the topic of future work.

\begin{figure}
  \centering
  \includegraphics[width=\columnwidth]{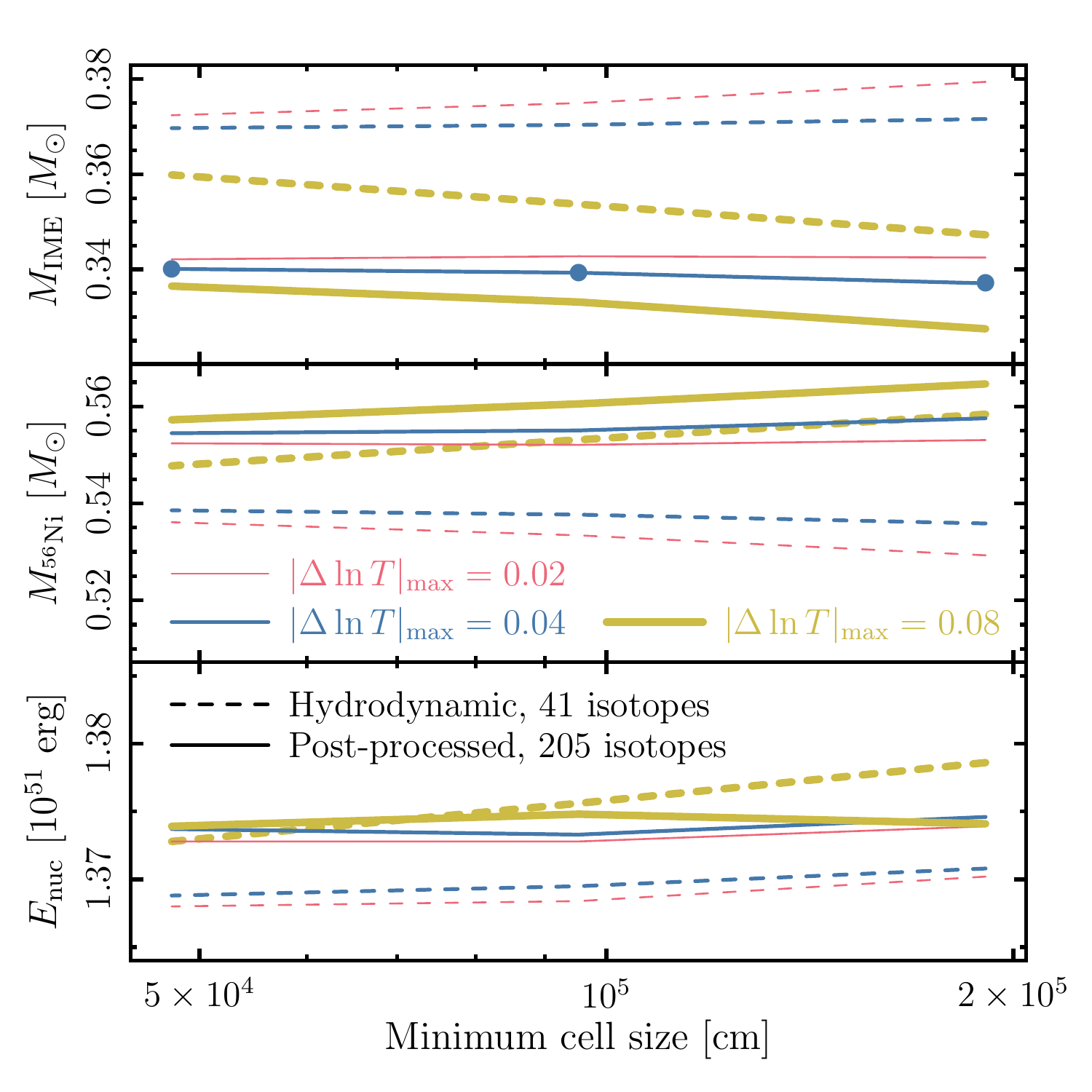}
  \caption{Synthesized IME (\emph{top panel}) and $^{56}$Ni (\emph{middle panel}) masses and total nuclear energy release (\emph{bottom panel}) vs.\ minimum cell size allowed in the simulation for  detonations of $1.0 \msol$ $50/50$ C/O solar metallicity WDs.  Dashed lines represent the results from our hydrodynamic simulations, which use a 41-isotope network, for a maximum relative temperature change per timestep of $ | \Delta \ln T |_{\rm max}=0.02$ (\emph{thin red dashed}), $0.04$ (\emph{medium blue dashed}), and $0.08$ (\emph{thick yellow dashed}).  Post-processed results using a $205$-isotope network are shown as solid lines for $ | \Delta \ln T |_{\rm max}=0.02$ (\emph{thin red solid}), $0.04$ (\emph{medium blue solid}), and $0.08$ (\emph{thick yellow solid}). Circles in the top panel show the minimum cell sizes of the convergence study for reference.}
  \label{fig:resolution}
\end{figure}

Figure \ref{fig:resolution} shows synthesized masses of $^{56}$Ni and intermediate-mass elements (IMEs; defined as having charges $11 \leq Z \leq 20$) and the total nuclear energy release, $E_{\rm nuc}$, for three sets of hydrodynamic simulations and three post-processed results vs.\ the minimum cell size in the simulation.  The dashed lines represent the hydrodynamic results, which use a 41-isotope network, and the solid lines show results from post-processing the same hydrodynamic simulations using a $205$-isotope network.

As Figure \ref{fig:resolution} demonstrates, global values are converged for minimum cell sizes $\lesssim  \unit[10^5]{cm}$ for both the $ | \Delta \ln T |_{\rm max}=0.02$ and $0.04$ cases, with relevant quantities changing by $< 1\%$ with a factor of two increase in resolution.  Results for the $ | \Delta \ln T |_{\rm max}=0.08$ simulation do not appear to be fully converged at our highest resolution, which motivates our choice of $ | \Delta \ln T |_{\rm max}=0.04$ for all of the production runs in this work.

The bulk nucleosynthetic yields of the hydrodynamic results without post-processing and the results after post-processing are discrepant at a $3-10\%$ level.  However, as previously discussed, the 41-isotope nuclear reaction network used in the hydrodynamics simulations is designed to capture energetics, not isotopic abundances.  Thus, the agreement in energetics before and after post-processing is much better, with only a $ \simeq 0.3\%$ difference.  This gives confidence that the tracer particles' density and temperature histories used in the post-processing calculation and the resulting nucleosynthetic abundances are accurate.


\subsection{Spatially resolved broadened detonation structure}
\label{sec:resolved}

\begin{figure}
  \centering
  \includegraphics[width=\columnwidth]{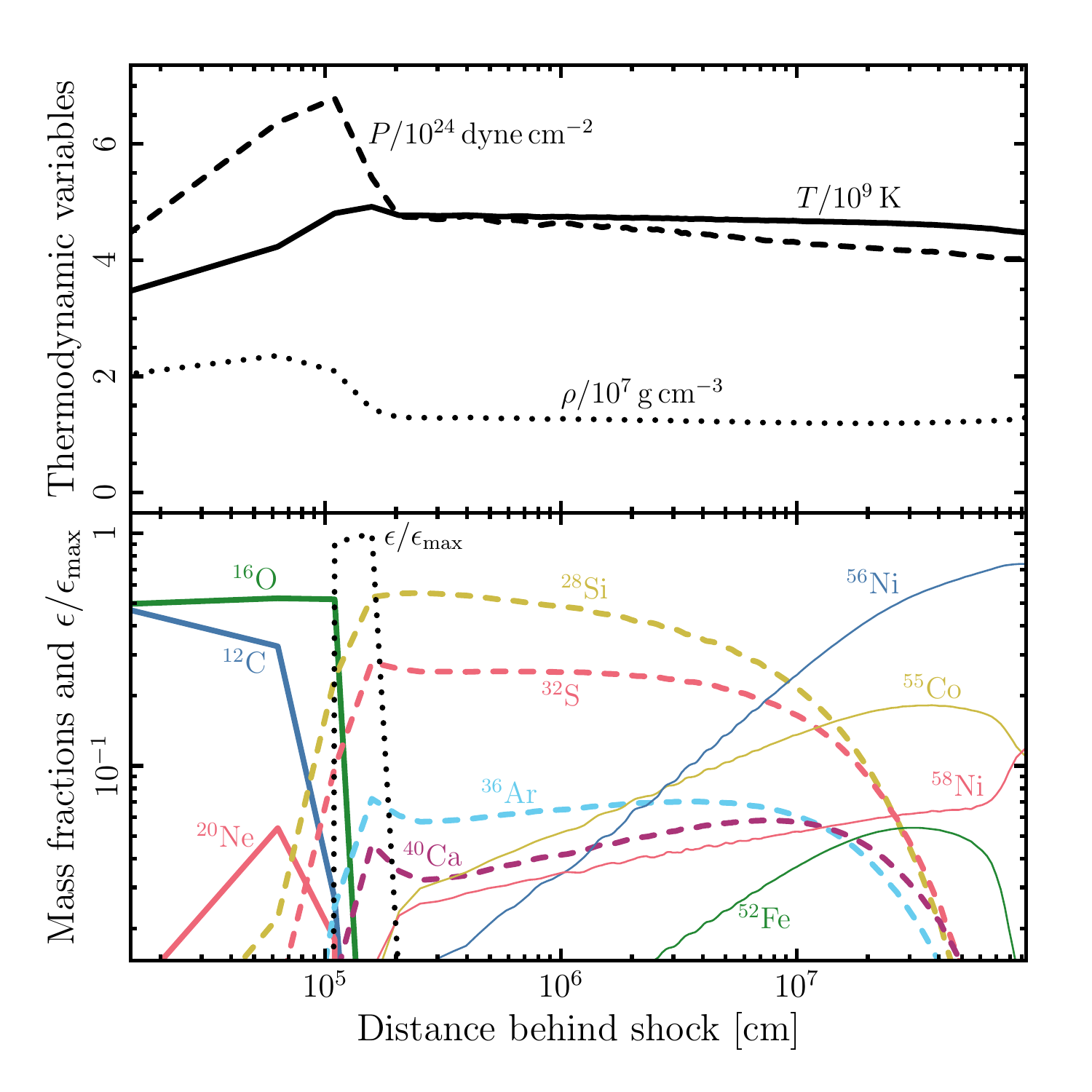}
  \caption{\emph{Top panel}: The pressure, temperature, and density, normalized as labeled, vs.\ distance behind the shock in a $1.0 \msol$ $50/50$ C/O WD detonation $\unit[0.24]{s}$ after the beginning of the simulation. \emph{Bottom panel}: Profiles of mass fractions and the normalized energy generation rate, $\epsilon/\epsilon_{\rm max}$. The other isotopes that do not appear in the panel do not reach mass fractions $>10^{-2}$ at this stage of the detonation.}
  \label{fig:lxvslr}
\end{figure}

Our burning limiter allows us to spatially resolve the artificially broadened detonation structure in our hydrodynamic simulations, an example of which is shown in Figure \ref{fig:lxvslr}.  The top panel shows thermodynamic variable profiles, and the bottom panel shows profiles of the energy generation rate normalized to the maximum value, $\epsilon / \epsilon_{\rm max}$, and the mass fractions of 11 isotopes as labeled.  The other 30 isotopes comprising the 41-isotope network used in our hydrodynamical simulations do not reach mass fractions above $10^{-2}$ in this plot at this time, $\unit[0.24]{s}$ after the simulation has begun.  The time of the snapshot is chosen to coincide with when the detonation reaches the mass coordinate ($0.64 \msol$ from the center) where the $^{56}$Ni fraction will equal the $^{28}$Si fraction after the simulation ends.

The density upstream of the detonation at this time is $\unit[6.3\E{6}]{g \, cm^{-3}}$.  The carbon consumption lengthscale for a steady state detonation at this density is $\sim \unit[10^2]{cm}$, and the lengthscale for an overdriven detonation such as this is even shorter \citep{khok89,town16a}.  Due to the use of our burning limiter, we achieve a spatially resolved detonation by construction.  The broadened detonation in our simulation has a carbon consumption lengthscale of $\sim \unit[10^5]{cm}$, $>10$ times longer than the true lengthscale, and the maximum of the energy generation rate is several zones behind the shock front instead of just behind or inside the shock where it would be located for an unresolved detonation.  While the detonation itself is not physically correct, the convergence study in \S \ref{sec:conv} gives us confidence that the major yields will be relatively unchanged at higher resolutions.  These yields will be verified by comparison to resolved calculations in future work.


\subsection{Ejecta profiles}
\label{sec:rhovsv}

\begin{figure}
  \centering
  \includegraphics[width=\columnwidth]{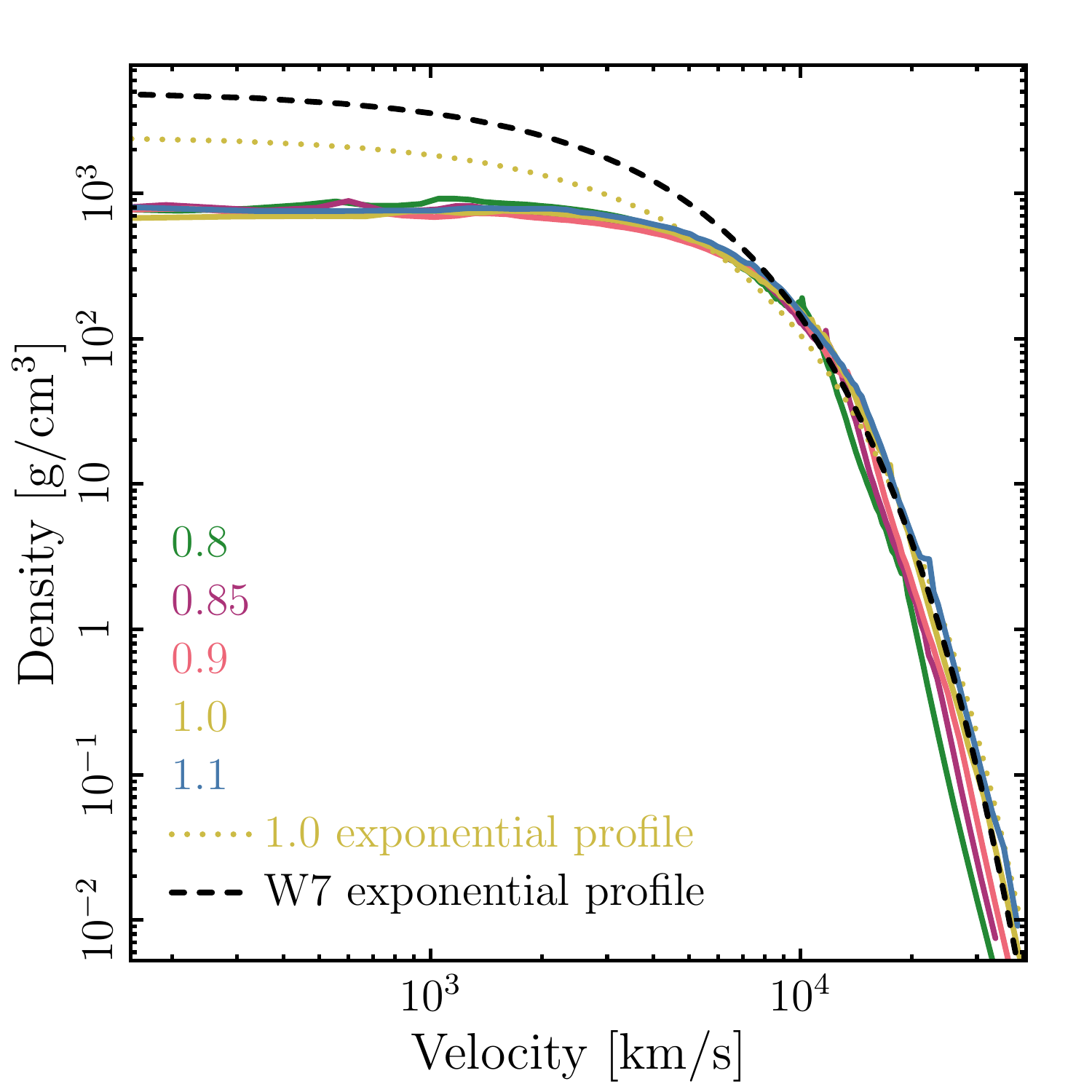}
  \caption{Density vs.\ velocity profiles $\unit[10]{s}$ after the beginning of the simulation.  Models for all five WD masses are shown as labeled for an initial $50/50$ C/O mass fraction.  Exponential parameterizations of our  $1.0 \msol$ model and \cite{nty84}'s W7 model are shown as yellow dotted and black dashed lines, respectively.}
  \label{fig:rhovsv}
\end{figure}

Figure \ref{fig:rhovsv} shows density vs.\ velocity profiles $\unit[10]{s}$ after the simulation begins for our five WD masses with an initial C/O mass fraction of $50/50$.  The profiles are all relatively similar: a nearly constant density core surrounded by an exponentially declining density beyond $\sim \unit[10^4]{km \, s^{-1}}$.

Studies of SN Ia ejecta interaction with surrounding material often use an exponential parameterized approximation to the ejecta profile (e.g., \citealt{dwar98a}):
\begin{equation}
	\rho(v,t) = \frac{6^{3/2}}{8 \pi} \frac{ M_e^{5/2}}{E_{\rm kin}^{3/2}} \frac{  \exp{ \left( -v/v_e \right) } }{t^3} ,
\end{equation}
where the kinetic energy is $E_{\rm kin}$, the ejecta mass is $M_e$, and $v_e = \left( E_{\rm kin} / 6 M_e \right)^{1/2}$.  We plot this parameterization for our $1.0 \msol$ model as a yellow dotted line.  We also plot the exponential parameterization of \cite{nty84}'s $M_{\rm Ch}$ W7 model as a dashed line for comparison.

In the outer regions $\geq \unit[10^4]{km \, s^{-1}}$, the exponential approximation provides a reasonable fit to our model.  However, in the inner $0.2 \msol$, the exponential parameterization of our model and of W7 yield substantially higher densities with a steeper slope than found in our simulations.  These differences will have a significant impact on modeling of the nebular and SN remnant phases, when these inner regions become optically thin.  Indeed, \cite{boty17a} have recently found better agreement with the nebular spectra of SN 2011fe when using parameterized ejecta profiles with constant density cores instead of exponential profiles.  Future modeling of nebular spectra and emission from SN remnants using the ejecta profiles from our hydrodynamic simulations will enable more quantitative comparisons to observations.


\section{Nucleosynthetic results}
\label{sec:results}

We now describe the nucleosynthetic products of our post-processed models.  After presenting the bulk yields and comparing them to previous work, we will discuss our trace abundances in the context of observations from late-time SN Ia light curves, the solar Mn abundance, and SN remnant abundances.  These observations constrain the amount of neutron-rich nucleosynthesis in SNe Ia, an important discriminant between $M_{\rm Ch}$ and sub-$M_{\rm Ch}$ progenitors.


\subsection{Bulk yields and comparison to literature}
\label{sec:bulk}

In this section, we report the yields of low-mass elements (LMEs; $Z \le 10$), IMEs, high-mass elements (HMEs; $21 \le Z$), and $^{56}$Ni and compare our results to previous work.


\subsubsection{Yield profiles and integrated masses}

\begin{figure}
  \centering
  \includegraphics[width=\columnwidth]{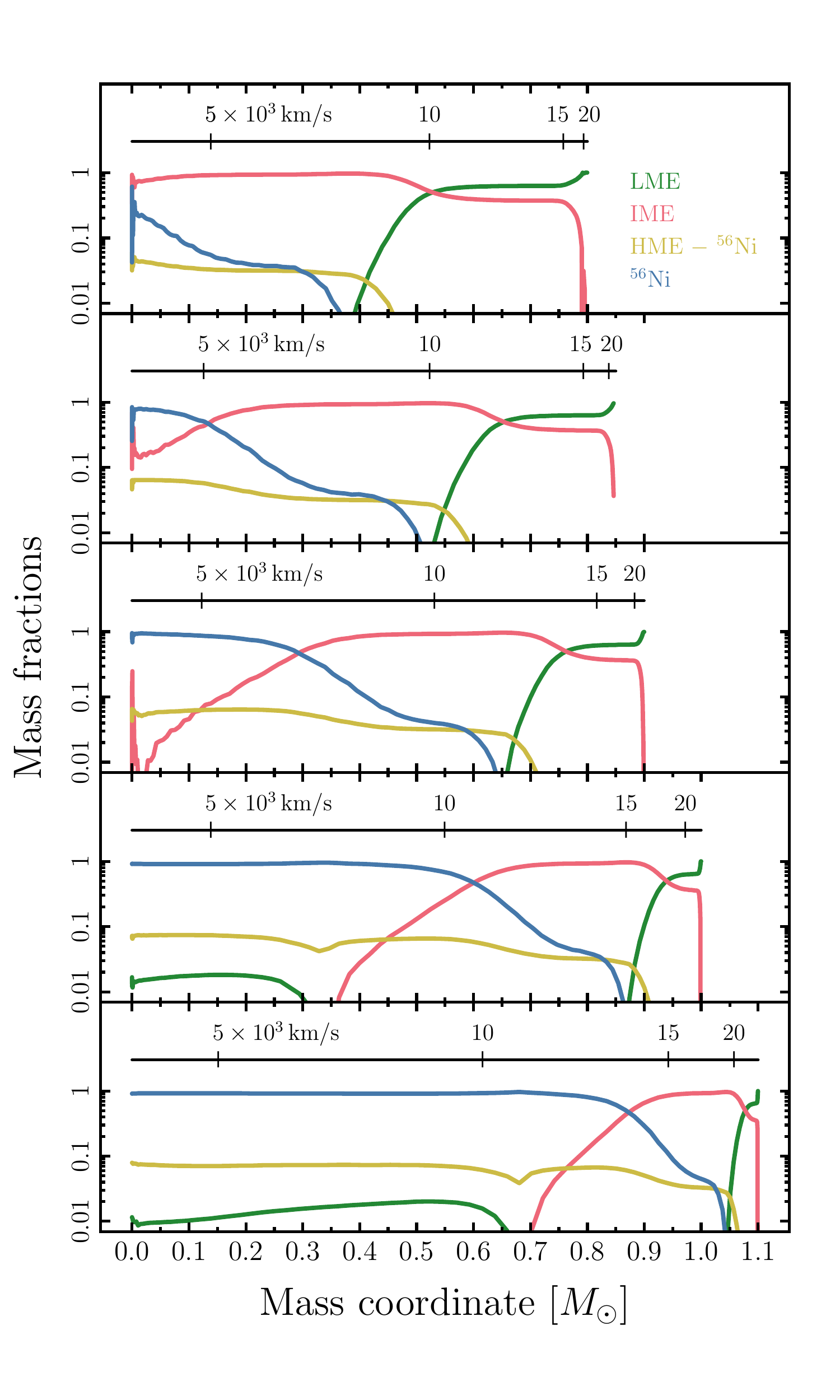}
  \caption{Mass fractions of LMEs (\emph{green}), IMEs (\emph{red}), HMEs excluding $^{56}$Ni (\emph{yellow}), and $^{56}$Ni (\emph{blue}) vs.\ mass coordinate.  The five panels show post-processed results for WD masses of $0.8-1.1 \msol$, from top to bottom.  The initial compositions of the simulations have C/O mass fractions of $50/50$ and solar metallicity.  The top bar in each panel shows the locations of velocities in increments of $\unit[5000]{km \, s^{-1}}$.}
  \label{fig:xvsm}
\end{figure}

In Figure \ref{fig:xvsm}, we show mass fractions of LMEs, IMEs, HMEs excluding $^{56}$Ni, and $^{56}$Ni vs.\ mass coordinate.  The five panels represent the post-processing results of different WD masses ($0.8-1.1 \msol$ from top to bottom) with initial compositions of $50/50$ C/O and solar metallicity.  Also marked are the mass coordinates of velocities in increments of $\unit[5000]{km \, s^{-1}}$.

The profiles show stratified composition structures as expected for one-dimensional pure detonations with no mixing.  $^{56}$Ni and other HMEs are produced in the center of the WDs and extend out to varying mass coordinates depending on the WD mass.  This material is surrounded by a layer of IMEs, which is in turn surrounded by a LME cap primarily composed of $^{16}$O.

One interesting feature is the presence of $^4$He with a mass fraction of $ \sim 0.01$ in the central few tenths of a solar mass of the more massive $1.0$ and $1.1 \msol$ WDs.  This is indicative of the $\alpha$-rich freezeout from nuclear statistical equilibrium (NSE) characteristic of nuclear burning at these temperatures and densities \citep{woos73a,seit13a}, which will have an effect on the production of neutron-rich isotopes discussed in \S \ref{sec:nrich}.  The presence of $^4$He in the core, mixed with $^{56}$Ni, could result in an interesting signature in late-time nebular spectra; we leave an analysis of its effect to future work.

\begin{figure}
  \centering
  \includegraphics[width=\columnwidth]{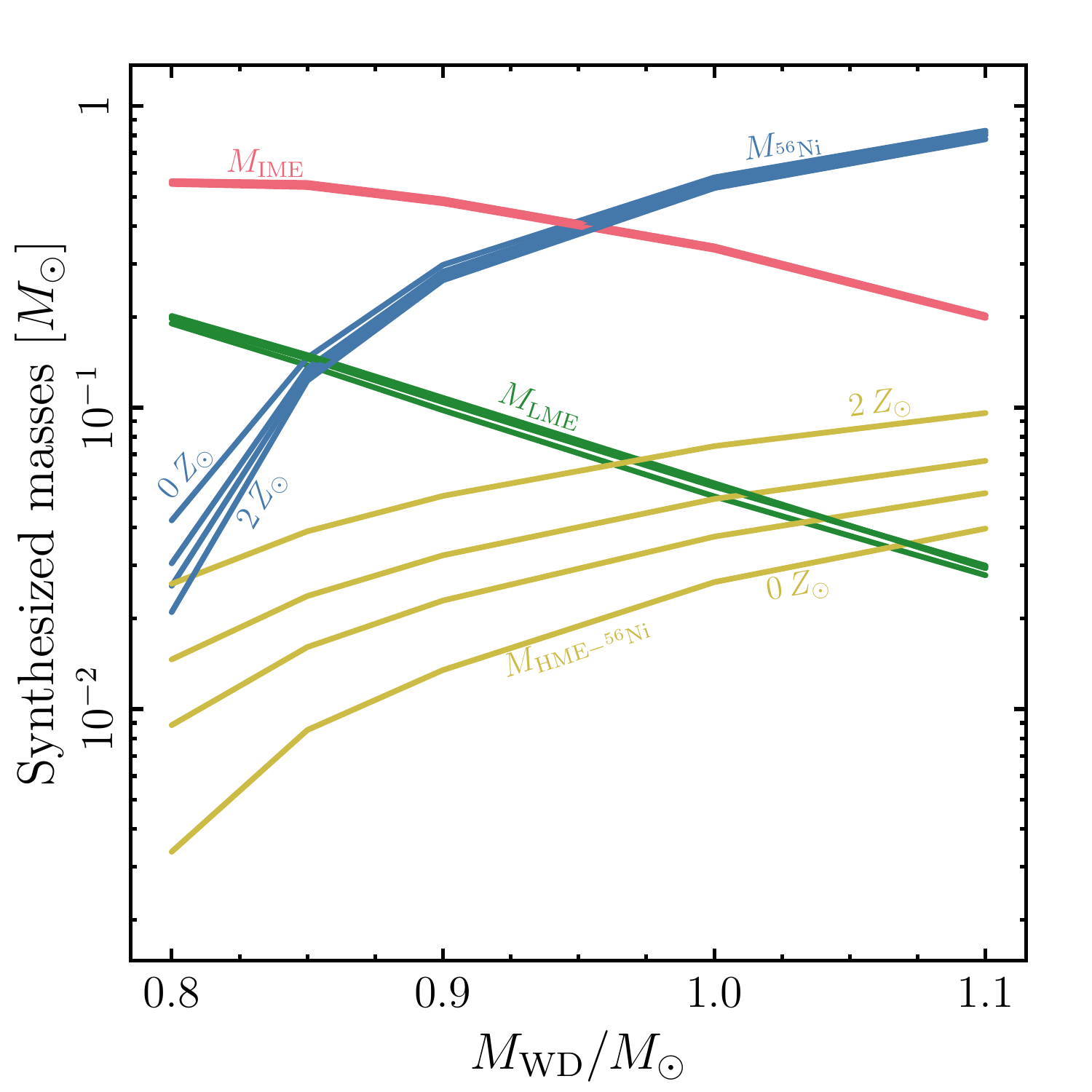}
  \caption{Bulk synthesized masses vs.\ WD mass.  Shown are LME (\emph{green}), IME (\emph{red}), non-$^{56}$Ni HME (\emph{yellow}), and $^{56}$Ni (\emph{blue}) masses for an initial C/O ratio of $50/50$ by mass.  Four metallicities for each C/O composition are shown: $0$, $0.5$, $1$, and $2 \, Z_\odot$.  Decreasing the metallicity decreases the non-$^{56}$Ni HME mass but increases the $^{56}$Ni mass while leaving the LME and IME masses relatively unchanged.}
  \label{fig:mvsm}
\end{figure}

Figure \ref{fig:mvsm} shows post-processed results for total synthesized masses vs.\ WD mass for an initial C/O mass fraction of $50/50$ and four initial metallicities.  Increasing the metallicity increases the non-$^{56}$Ni HME mass but decreases the $^{56}$Ni mass; the LME and IME masses are relatively constant with respect to the metallicity.  The $^{56}$Ni dependence on the metallicity for our high-mass models is similar to that found for $M_{\rm Ch}$ explosions \citep{tbt03}.  We obtain a $\sim 10\%$ decrease in $^{56}$Ni mass for a $1.0 \msol$ WD detonation when the initial metallicity is changed from $0$ to $2 \, Z_\odot$.  However, there is a more drastic dependence for the low-mass models: a zero metallicity $0.8 \msol$ WD detonation produces almost a factor of two more $^{56}$Ni than a $2 \, Z_\odot$ explosion.


\subsubsection{Comparison of bulk yields to other results}
\label{sec:bulkcomp}

\begin{figure}
  \centering
  \includegraphics[width=\columnwidth]{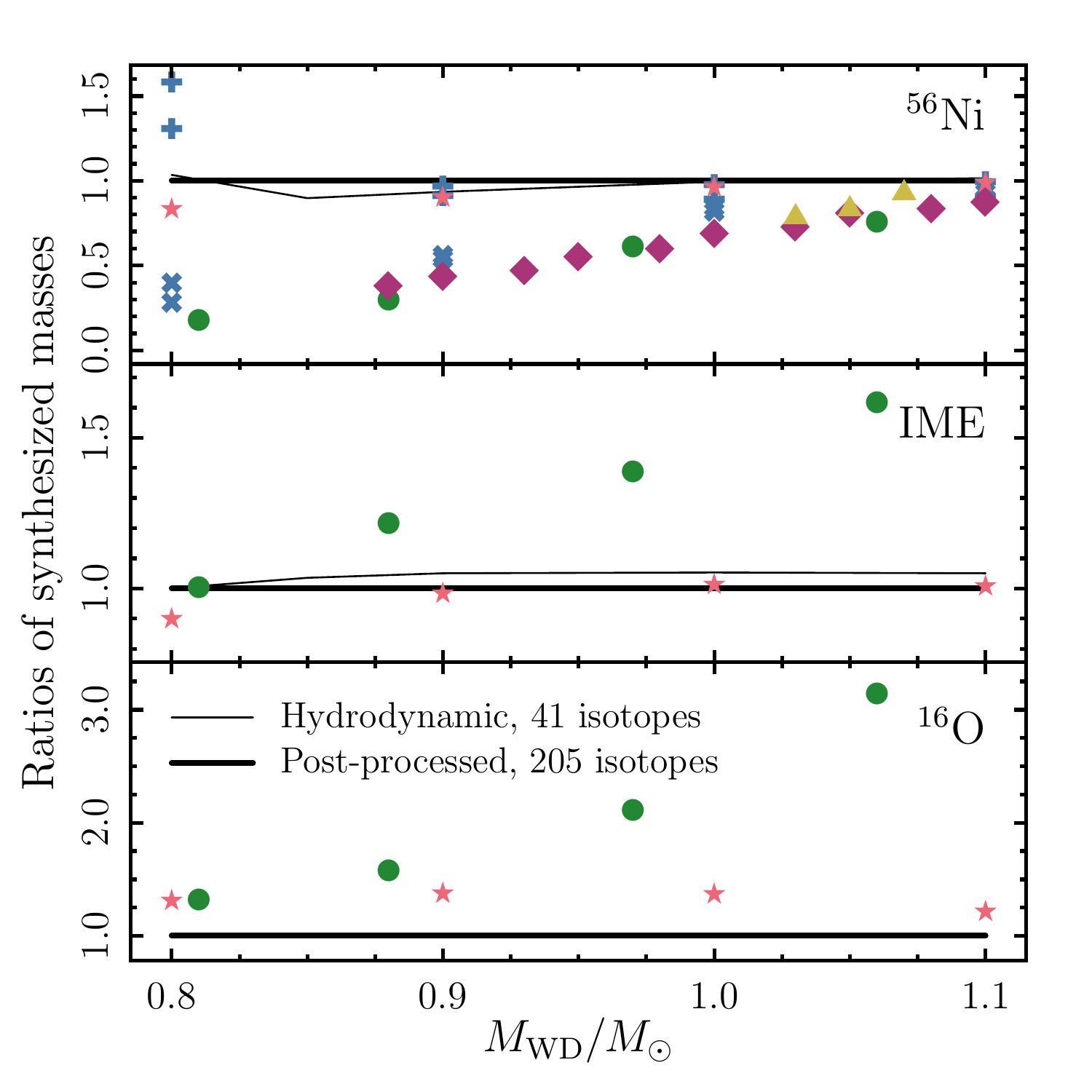}
  \caption{Ratios of synthesized $^{56}$Ni (\emph{top panel}), IME (\emph{middle panel}), and $^{16}$O masses (\emph{bottom panel}) to post-processed masses vs.\ WD mass.  Thin and thick lines represent our 41-isotope hydrodynamic results before post-processing and our 205-isotope post-processed results, respectively, for an initial C/O ratio of $50/50$ and zero metallicity.  Symbols show results from other studies: \citet[\emph{green circles}]{sim10}, \citet[\emph{yellow triangles}]{shig92}, \citet[\emph{blue crosses and plus signs}]{moll14a}, \citet[\emph{magenta diamonds}]{blon17a}, and results using the method described in \citet[\emph{red stars}]{town16a}.  Post-processed models with the appropriate metallicities are used to calculate these ratios.}
  \label{fig:mvsm_comp}
\end{figure}

In Figure \ref{fig:mvsm_comp}, we show a comparison of our hydrodynamic and post-processed bulk yields to previous work.  The top, middle, and bottom panels show the ratios of total synthesized masses of $^{56}$Ni, IMEs, and $^{16}$O, respectively, to our post-processed results.  Our hydrodynamic results for an initial composition of $50/50$ C/O and zero metallicity are shown as thin lines, and the post-processed results are shown as thick lines.  Green circles represent zero metallicity synthesized masses from \cite{sim10}, yellow triangles demarcate \cite{shig92}'s $^{56}$Ni masses with an initial metallicity of $\sim 2 \, Z_\odot$, blue crosses and plus signs are zero metallicity $^{56}$Ni masses resulting from 19-isotope and 199-isotope simulations by \cite{moll14a}, and magenta diamonds are $\sim Z_\odot$ results from \cite{blon17a}, respectively.  The ratios are calculated using our post-processed yields from models with the appropriate initial metallicity.

Red stars represent zero metallicity results from a parameterized model for burning in \FLASH\ \citep{cald07a,town07,town09a,town16a}, in which the detonation front is tracked by progress variables that measure the fractions of fuel, ash, quasi-NSE material, and NSE material.  This front tracking scheme is used in a hydrodynamic \FLASH\ simulation with minimum cell size of $\unit[1.25\E{4}]{cm}$ and zero metallicity, whose results are then post-processed with the same $205$-isotope network used throughout the rest of this work.  A similar procedure was also used in \cite{mart17a}.

The results of our hydrodynamic and post-processed burning limiter simulations are very similar to the parameterized model results using progress variables, which has been verified against resolved calculations of planar steady-state detonations \citep{town16a}, giving us further confidence that our results are converged.  The burning in both methods is systematically more complete (e.g., more $^{56}$Ni is produced) than all of the other studies except for the large network results of \cite{moll14a} at low WD masses $\leq 0.9 \msol$.  For a WD mass of $0.9$ $(1.0) \msol$, our post-processed model yields a $^{56}$Ni mass of $0.30$ $(0.58) \msol$, while a quadratic fit to \cite{sim10}'s results implies a mass of $0.11$ $(0.38) \msol$.  These abundance differences will be reflected in our radiative transfer predictions (\S \ref{sec:rad}), enabling typical SNe Ia to be produced by $1.0 \msol$ WDs instead of $1.1 \msol$ WDs as found by \cite{sim10}.  This will imply, among other things, a higher predicted rate of SNe Ia because less massive WDs are more numerous.  It is also apparent that the total mass burned in \cite{sim10}'s simulations is more steeply dependent on WD mass than we have found.  This likely contributes to the difference in the slope of the brightness-decline rate relation that we show in \S \ref{sec:rad}.

It is unclear why \cite{sim10}'s nucleosynthetic results differ so significantly from ours.  We note that \cite{sim10}'s $^{56}$Ni masses are in rough agreement with those of \cite{shig92} in their limited mass range (yellow triangles in the top panel of Fig.\ \ref{fig:mvsm_comp}), especially after adjusting for \cite{sim10}'s initial composition of zero metallicity and \cite{shig92}'s $\sim 2 \, Z_\odot$ initial composition.  However, possibly due to a neglect of Coulomb corrections, the central densities reported by \cite{shig92} are systematically lower than we, \cite{sim10}, and others calculate, and thus their derived $^{56}$Ni masses will also be lower.  Therefore, \cite{sim10}'s agreement with \cite{shig92} is consistent with both of their reported  $^{56}$Ni masses being too low.  

The discrepancy between our results and those of \cite{blon17a}, and to a lesser extent the $19$-isotope calculations of \cite{moll14a}, is easier to explain.  Smaller networks may neglect burning pathways that become increasingly important for lower density, low mass WDs.  This is particularly true for the 4-stage network used by \cite{blon17a}.  The discrepancy is less severe for higher WD masses because much of the IGE nucleosynthesis occurs in NSE, which erases details of the nuclear reaction network and the detonation structure.  However, their $0.88 \msol$ model produces just $1/3$ of the $^{56}$Ni that our calculations imply.  Such a large difference in $^{56}$Ni abundance will have a significant impact on radiative transfer calculations, particularly for subluminous SNe Ia, an effect we will discuss in more detail in \S \ref{sec:rad}.


\subsection{Neutron-rich nucleosynthesis}
\label{sec:nrich}

While simulations of deflagrations, detonations, and deflagration-to-detonation transitions of C/O WDs generally produce similar bulk nucleosynthetic results at the order of magnitude level, the different explosion mechanisms yield large differences in trace abundances.  This is especially true for neutron-rich isotopes.  The higher densities and longer timescales involved in $M_{\rm Ch}$ deflagration-to-detonation transition explosions allow for weak reactions that can significantly reduce the electron fraction from its initial value close to $0.5$.  Some neutron-rich isotopes are produced in our pure detonation simulations, particularly in regions that undergo incomplete silicon-burning, but the overall abundances are lower due to the $\alpha$-rich freezeout from NSE that occurs in the core.

Some models of nebular spectra have implied the production of up to $0.2 \msol$ of neutron-rich stable IGEs in the center of SN Ia ejecta (e.g., \citealt{mazz07,mazz15a}).  However, there is some disagreement about the required amount of stable IGEs, in part due to uncertainties in the ejecta density profile (\S \ref{sec:rhovsv}).  \cite{liu97a} found that the sub-$M_{\rm Ch}$ double detonation model of \cite{ww94} with $0.02 \msol$ of stable IGEs provides the best density and composition profile for a nebular spectrum of SN 1994D.  More recently, \cite{boty17a} arrive at the conclusion that a stable IGE core is not required to match the nebular spectra of SN 2011fe and may in fact be disfavored.

Deriving the amount of stable IGE from nebular spectra is complicated by the fact that if there is a surviving WD companion, it will capture some $^{56}$Ni from the SN ejecta \citep{shen17a}.  Some of this accreted $^{56}$Ni will be hot enough to be fully ionized and will have a slower rate of decay due to its inability to capture electrons.  Thus, there may be an additional source of heating that is currently unaccounted for in nebular phase studies, which will change the masses inferred from observations.

We leave a detailed study of the nebular spectra expected from our pure detonation models to future work.  In the following sections, we explore other probes of neutron-rich nucleosynthesis: late-time light curve observations, the solar abundance of Mn, and abundance estimates from SN remnant observations.


\subsubsection{Late-time light curve observations}
\label{sec:latetime}

Several of the neutron-rich isotopes produced in SNe Ia have a significant impact on the late-time light curves after $\unit[800]{d}$.  At these late phases, $\gamma$-ray trapping is inefficient, and the predominant energy source is the thermalization of positron and electron kinetic energy.  These leptons arise from the decay of $^{56}$Co (half-life of $\unit[77]{d}$, produced primarily as $^{56}$Ni) and the neutron-rich isotopes $^{57}$Co (half-life of $\unit[272]{d}$, produced primarily as $^{57}$Ni) and $^{55}$Fe (half-life of $\unit[1000]{d}$, produced primarily as $^{55}$Co) \citep{seit09b,roep12a}.

Several recent nearby SNe Ia (SN 2011fe, SN 2012cg, and SN 2014J) have been observed to late enough phases to estimate the abundances of these neutron-rich isotopes from their contribution to the light curve.  The implied mass ratio of $^{57}$Co to $^{56}$Co at these late times ranges from $0.02$ to $0.09$ \citep{grau16a,dimi17a,shap17a,yang18a}, while the $^{55}$Fe to $^{57}$Co mass ratio has been estimated to be $<0.2$ \citep{shap17a}, albeit with large error bars.

\begin{figure}
  \centering
  \includegraphics[width=\columnwidth]{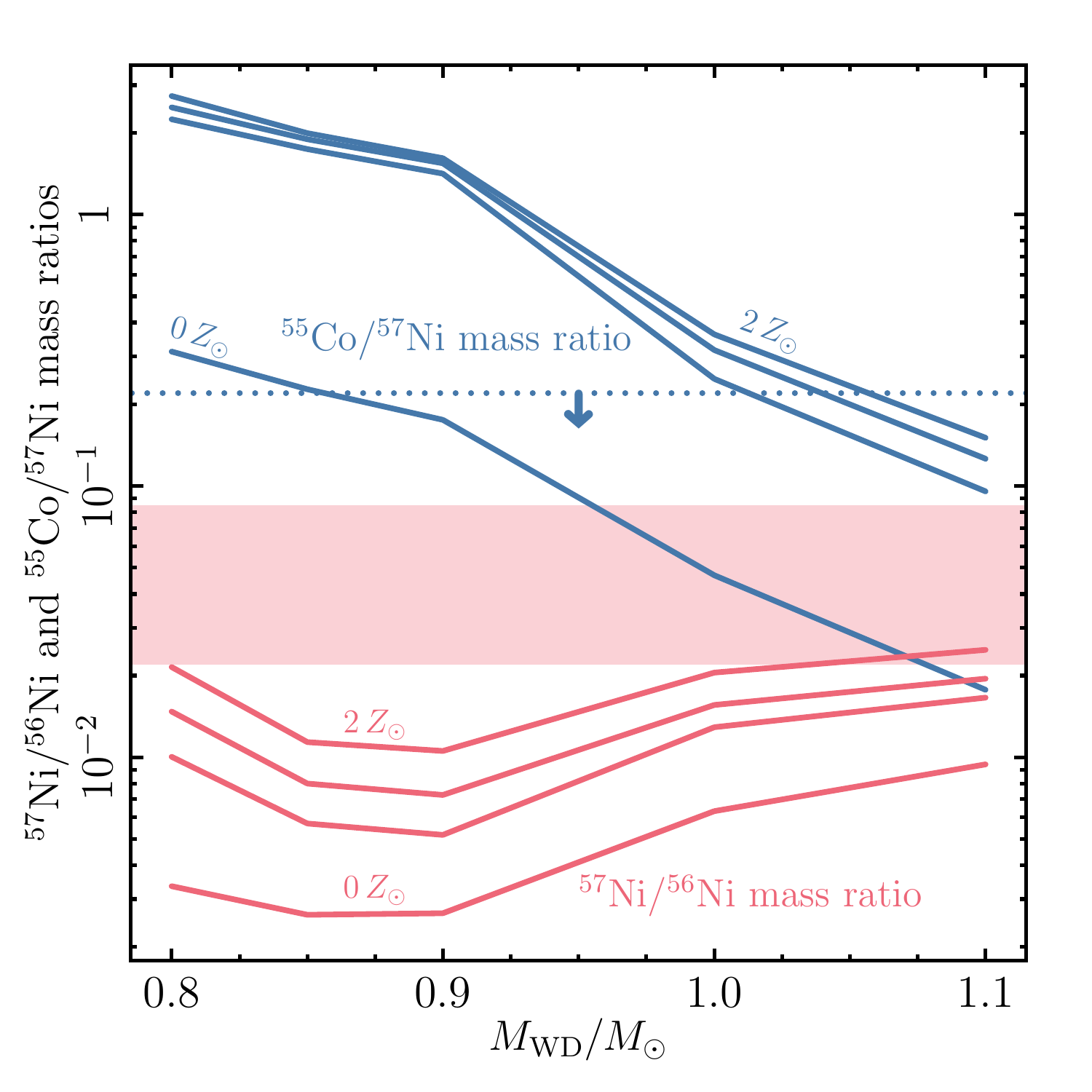}
  \caption{Mass ratios of $^{57}$Ni to $^{56}$Ni (\emph{red}) and $^{55}$Co to $^{57}$Ni (\emph{blue}) vs.\ WD mass from our post-processed nucleosynthetic results for an initial C/O ratio of $50/50$.  Four metallicities are shown, increasing from bottom to top  for each mass ratio: $0$, $0.5$, $1$, and $2 \, Z_\odot$.  The blue dotted line shows an upper limit to the $^{55}$Co to $^{57}$Ni ratio in SN 2011fe \citep{shap17a}, and the red shaded region shows a range of estimated $^{57}$Ni to $^{56}$Ni ratios for SN 2011fe \citep{dimi17a,shap17a}, SN 2012cg \citep{grau16a}, and SN 2014J \citep{yang18a}.}
  \label{fig:nrich57}
\end{figure}

In Figure \ref{fig:nrich57}, we show the mass ratios of $^{57}$Ni to $^{56}$Ni and $^{55}$Co to $^{57}$Ni  produced in our explosions for a range of metallicities.   The initial C/O ratio for all models is $50/50$.  Changing the initial C/O ratio alters the mass ratios at a minimal level; we do not plot these results for simplicity. The upper limit to the $^{55}$Co to $^{57}$Ni mass ratio from \cite{shap17a} is shown as a blue dotted line, and the range of  $^{57}$Ni to $^{56}$Ni ratios inferred from observations is shown as a red shaded region.

The increase in the $^{57}$Ni/$^{56}$Ni ratio with mass for masses $ \geq 0.9 \msol$ is due to the changing detonation regimes: as the WD mass increases, the primary mode of burning transitions from incomplete silicon-burning to an $\alpha$-rich freezeout from NSE, with an accompanying change in the $^{57}$Ni/$^{56}$Ni ratio \citep{woos73a}.  However, the reason for the decrease in the ratio with increasing mass below $0.9 \msol$ is uncertain.  Similarly, the origin of the large gap in the $^{55}$Co/$^{57}$Ni ratio between zero and half solar metallicity models is unknown.  This gap is driven by the metallicity dependence of the $^{55}$Co yield, which is also displayed in Figure  \ref{fig:nrich}, but the reason for this dependence is unclear.  We leave exploration of these trends to future work.

Our $1.1 \msol$ results agree broadly with \cite{pakm12b}'s values for a $0.9+1.1 \msol$ violent merger of two WDs, whose nucleosynthesis is primarily determined by the explosion of the more massive WD.  Our results for the range of masses and metallicities do not alter the tension between the low $^{57}$Ni masses produced in sub-$M_{\rm Ch}$ detonation models and the higher masses inferred from late-time observations.  However, the  $^{57}$Ni and $^{55}$Co masses derived from observations have very large error bars due to the possible contribution of light echoes and uncertainties in the $\gamma$-ray and lepton trapping efficiencies.

Furthermore, the possibility of a surviving companion WD that complicates nebular spectra modeling will also have an influence here \citep{shen17a}.  If a companion WD survives the SN Ia explosion, it will  capture a small amount of $^{56}$Ni.  The radioactive decay of this accreted ejecta will be delayed due to the fully ionized $^{56}$Ni's inability to capture electrons, and so the surviving companion WD can supplement the SN Ia ejecta's late-time luminosity.  This additional luminosity will reduce the  amount of $^{57}$Co inferred from observations and possibly bring our nucleosynthetic results into agreement.  Ongoing and future late-time observations, particularly of SN 2011fe and SN 2014J, will shed further light on this issue; for now, we do not regard this tension as strong evidence against sub-$M_{\rm Ch}$ detonation models.


\subsubsection{Solar abundance of manganese}
\label{sec:solarMn}

\begin{figure}
  \centering
  \includegraphics[width=\columnwidth]{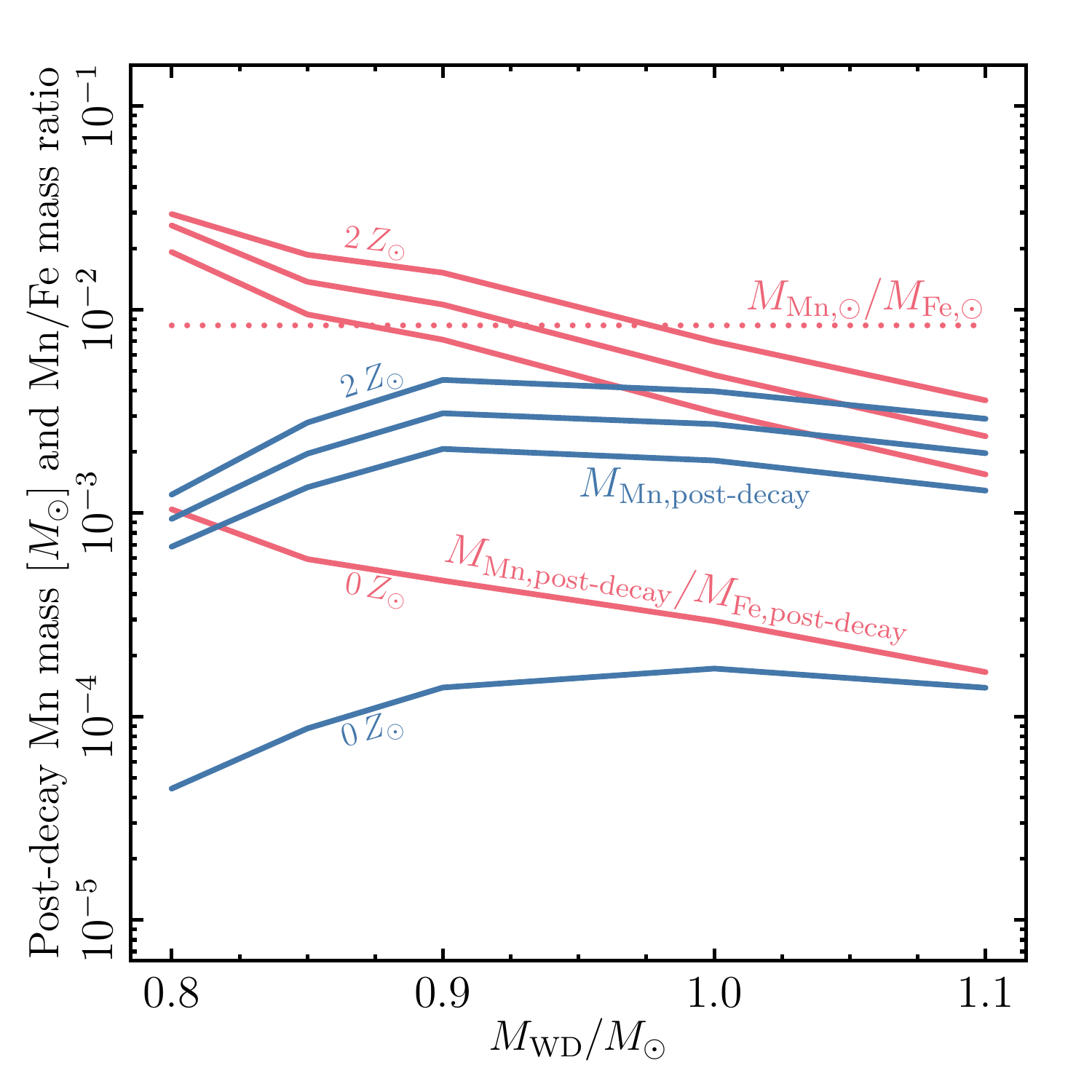}
  \caption{Mn mass (\emph{blue}) and Mn/Fe mass ratio (\emph{red}) after all radioactive decays have taken place vs.\ WD mass from our post-processed nucleosynthetic results for an initial C/O mass fraction of $50/50$.  Four metallicities increasing from bottom to top are shown: $0$, $0.5$, $1$, and $2 \, Z_\odot$.  The red dotted line shows the solar value \citep{aspl09a}.}
  \label{fig:nrich}
\end{figure}

The production, and subsequent decay, of the neutron-rich isotope $^{55}$Fe in SNe Ia contributes to the late-time luminosity, as described in the previous section, and is also the primary source of $^{55}$Mn in the Sun.  \cite{seit13a} argue that the known non-SN Ia sites of nucleosynthesis produce a sub-solar ratio of Mn to Fe after all relevant radioactive decays have occurred, and thus SNe Ia must make up the difference by producing a super-solar Mn/Fe ratio.  Because their representative sub-$M_{\rm Ch}$ model (a $0.9+1.1 \msol$ violent merger of two WDs; \citealt{pakm12b}) has a sub-solar Mn/Fe ratio, \cite{seit13a} conclude that $\sim 50\%$ of SNe Ia must occur via a deflagration-to-detonation transition explosion in a $M_{\rm Ch}$ WD.

In Figure \ref{fig:nrich}, we show the $^{55}$Mn mass produced in our post-processed simulations, after accounting for all radioactive decays, vs.\ WD mass for an initial C/O mass ratio of $50/50$.  Red lines show the mass ratio of Mn to Fe, again after all decays have occurred. As before, changing the initial C/O ratio has a minimal effect on  these results, so these models are omitted for simplicity.

For our lowest mass models, $^{55}$Co, which eventually decays to $^{55}$Mn, is produced via incomplete silicon-burning.  As the WD mass and central density are increased, more of the WD core undergoes incomplete silicon-burning, so the final $^{55}$Mn yield increases.  However, the $^{56}$Ni yield increases more strongly with WD mass, so the overall final Mn/Fe ratio decreases.  As the WD mass increases past $\sim 0.9 \msol$, some of the detonated material enters the regime of $\alpha$-rich freezeout from NSE, which reduces the yield of $^{55}$Co \citep{woos73a,seit13a} and the Mn/Fe ratio.  Presumably, for even higher mass pure detonation models, the detonated material will reach conditions for a ``normal'' freezeout from NSE, and the $^{55}$Co yield will again increase with mass, but our highest mass WD explosions are not yet in this regime.

The solar value of the Mn/Fe mass ratio is shown as a red dotted line \citep{aspl09a}.  In agreement with \cite{pakm12b}'s $1.1 \msol$ WD detonation, our higher-mass $\ge 1.0 \msol$ models yield sub-solar Mn/Fe mass ratios.  However, a super-solar value is achieved for lower-mass $\le 0.9 \msol$ detonations at an initial metallicity of $0.5 \, Z_\odot$.

Thus, at least part of the discrepancy  found by \cite{seit13a} between the solar Mn/Fe ratio and nucleosynthesis in  sub-$M_{\rm Ch}$ detonations can be alleviated by including pure detonations of lower-mass WDs.  However, unless lower-mass detonations significantly outnumber higher-mass explosions, it is not clear that only including core collapse SNe and sub-$M_{\rm Ch}$ WD detonations will yield the solar Mn/Fe value.  The possibility remains that a combination of core collapse SNe, sub-$M_{\rm Ch}$ WD detonations, and the class of peculiar Type Iax SNe \citep{fole13a,fink14a} may yield the correct solar value, or that $M_{\rm Ch}$ explosions do indeed contribute to Mn production but at a lower fraction of all SNe Ia; further work is still required to solve this issue.


\subsubsection{SN remnant observations: Mn/Fe vs.\ Ni/Fe}
\label{sec:SNR_MnFeNiFe}

SN remnants serve as another probe of detailed nucleosynthesis in SN Ia explosions.  As the ejecta sweeps up the surrounding interstellar medium, a reverse shock propagates into the ejecta, exciting it to X-ray-emitting temperatures.  The resulting emission can be used to infer nucleosynthetic yields, although the process is complicated by noisy spectra, non-equilibrium ionization effects, asymmetric and inhomogeneous density distributions, and incomplete propagation of the reverse shock into the ejecta \citep{bade06a,vink12a}.

\begin{figure}
  \centering
  \includegraphics[width=\columnwidth]{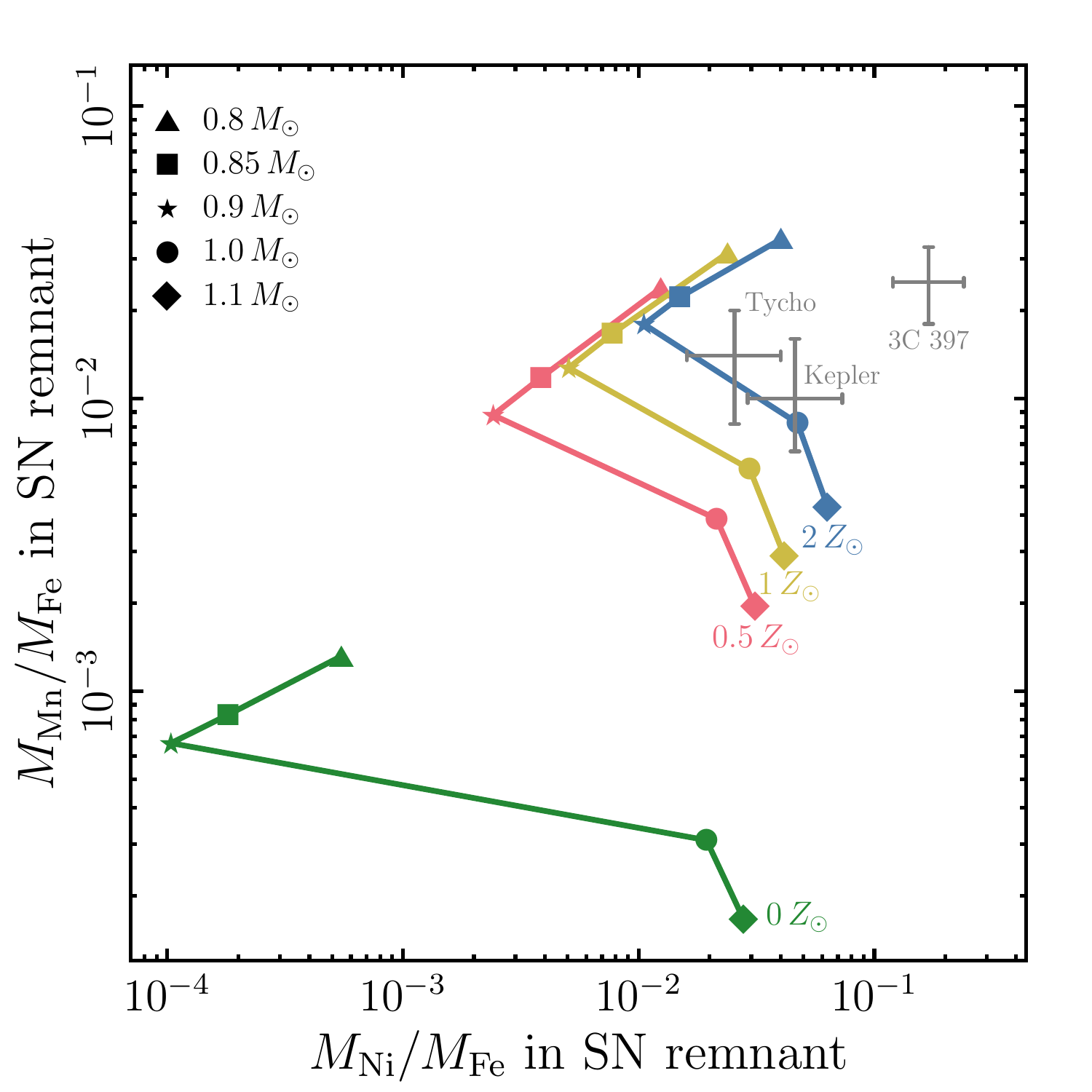}
  \caption{Mn/Fe vs.\ Ni/Fe mass ratios during the SN remnant phase, from our post-processed nucleosynthetic results for an initial C/O mass fraction of $50/50$.  WD masses of $0.8$ (\emph{triangles}), $0.85$ (\emph{squares}), $0.9$ (\emph{stars}), $1.0$ (\emph{circles}), and $1.1 \msol$ (\emph{diamonds}) are shown for four different metallicities: $0$ (\emph{green}), $0.5$ (\emph{red}), $1$ (\emph{yellow}), and $2 \, Z_\odot$ (\emph{blue}).  Gray error bars are observed values from \cite{yama15b}.}
  \label{fig:mnfenife}
\end{figure}

One such probe of SN Ia combustion conditions, the mass ratios of Mn/Fe and Ni/Fe, was examined by \cite{yama15b}.  In Figure \ref{fig:mnfenife}, we compare our post-processed nucleosynthetic mass ratios of Mn/Fe vs.\ Ni/Fe to their observational results, shown as gray symbols.  Five WD masses at  four initial metallicities  are shown for an initial C/O mass fraction of $50/50$.

Our ratios are calculated during the SN remnant phase, which for practical purposes we take to be between $\unit[10^2 - 10^5]{yr}$.  We thus account for isotopes that are present during this phase but  ultimately decay to another element.  For example, the Mn present during the SN remnant phase is predominantly the stable isotope $^{55}$Mn, but there is a small contribution from $^{53}$Mn, which decays to $^{53}$Cr with a half-life of $\unit[4\E{6}]{yr}$.  Thus, the Mn masses in Figures \ref{fig:nrich} and \ref{fig:mnfenife} differ slightly.  Likewise, the Ni present during the SN remnant phase is dominated by the stable isotopes $^{58}$Ni, $^{60}$Ni, and $^{62}$Ni, but the isotope $^{59}$Ni, with a half-life of $\unit[8\E{4}]{yr}$ can contribute a few percent by mass.

Our results are consistent with the sub-$M_{\rm Ch}$ detonation results calculated in \cite{yama15b}.  Thus, we also agree that matching the Tycho and Kepler SN remnant compositions requires somewhat super-solar metallicities, and that the composition of 3C 397 implies an unrealistically high initial metallicity if it was the product of a sub-$M_{\rm Ch}$ explosion.  \cite{yama15b} claim that this mismatch is evidence for a $M_{\rm Ch}$ explosion, but we emphasize that a $M_{\rm Ch}$ explanation also requires an extremely high metallicity, a complicated ejecta geometry, an unexpectedly high central density \citep{dave17a}, or a combination of all three.  Thus, the abundances in SN remnant 3C 397 continue to present a nucleosynthetic puzzle for any standard scenario.

\begin{figure}
  \centering
  \includegraphics[width=\columnwidth]{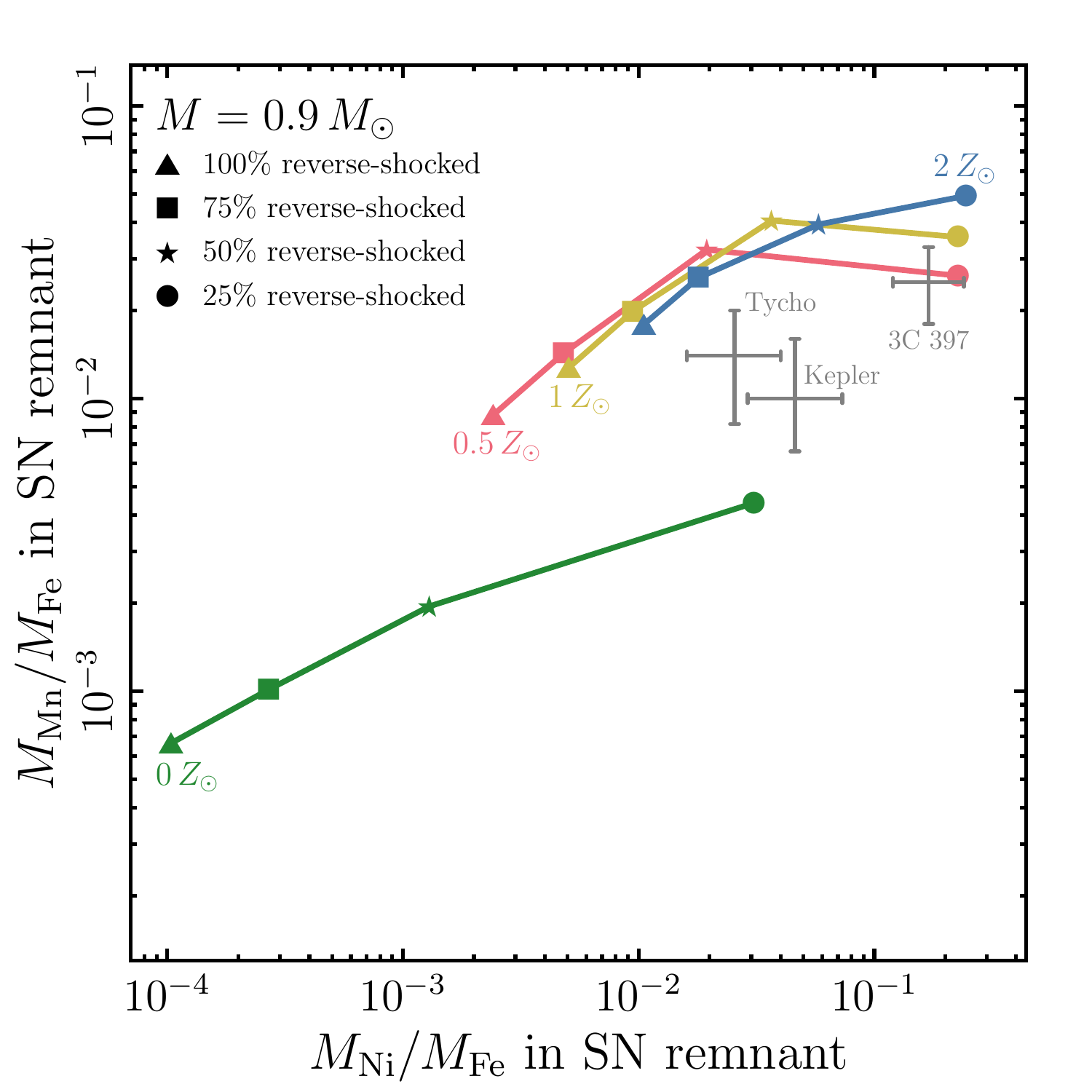}
  \caption{Mn/Fe mass ratio vs.\ Ni/Fe mass ratio for our $0.9 \msol$ models with varying metallicities and varying amounts of reverse-shocked ejecta.  Green, red, yellow, and blue curves represent models with initial metallicities of $0$, $0.5$, $1.0$, and $2.0 \, Z_\odot$, respectively.  The fraction of the ejecta that has been reverse-shocked decreases from 100\% on the left (\emph{triangles}) to 25\% on the right (\emph{circles}).}
  \label{fig:mnfenife_frac}
\end{figure}

The implication that Tycho and Kepler's exploding WDs had super-solar metallicities is also somewhat problematic, given the solar or slightly sub-solar metallicities of the stellar environments at their Galactocentric radii \citep{mart17a}.  However, this discrepancy can be at least partially explained by the fact that these remnants are young and their reverse shocks have not fully traversed the SN ejecta.  Thus, the inferred mass ratios may not be representative of the ejecta's total nucleosynthesis.

In Figure \ref{fig:mnfenife_frac}, we show Mn/Fe vs.\ Ni/Fe mass ratios for our $0.9 \msol$ models for a range of reverse-shocked ejecta fractions.  3C 397's SN remnant is likely fully reverse-shocked, so this analysis does not apply to it.  However, Figure \ref{fig:mnfenife_frac} shows that Tycho and Kepler's SNe may be explained as the explosions of $\sim 0.25 \, Z_\odot$ sub-$M_{\rm Ch}$ WDs with young remnants whose reverse shocks have only encountered $25-50\%$ of the total ejecta.  Given the Galactic positions of the SNe, the implication of highly sub-solar metallicity progenitors is not any more reasonable than the super-solar metallicities inferred from Figure \ref{fig:mnfenife}.  However, this analysis demonstrates the difficulty of ruling out progenitor models for young SN remnants using this particular diagnostic.


\subsubsection{SN remnant observations: Cr/Fe vs.\ Ca/S}
\label{sec:SNR_CrFeCaS}

\begin{figure}
  \centering
  \includegraphics[width=\columnwidth]{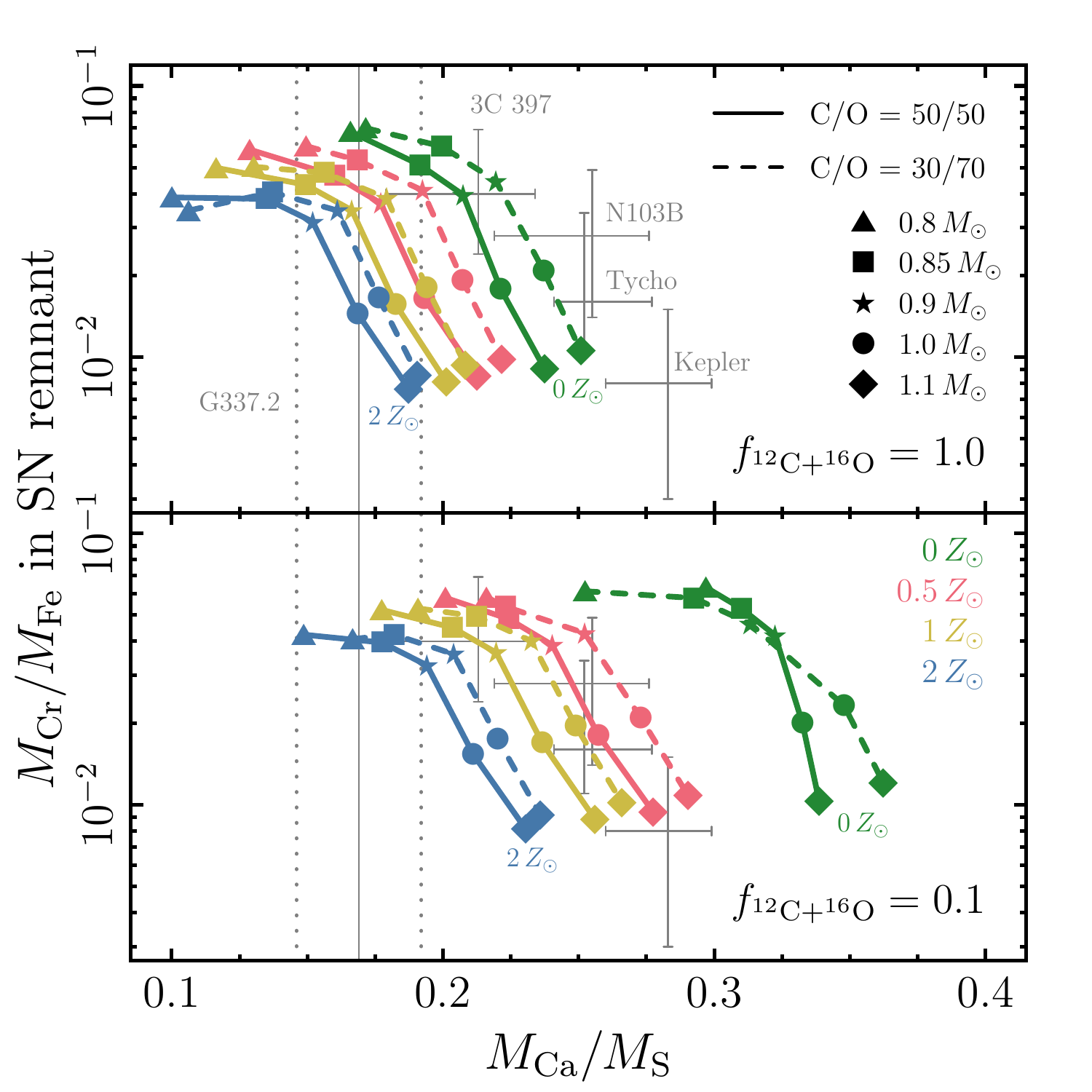}
  \caption{Cr/Fe vs.\ Ca/S mass ratios during the SN remnant phase.  Solid and dashed lines connect models with initial C/O mass fractions of $50/50$ and $30/70$, respectively.  WD masses of $0.8$, $0.85$, $0.9$, $1.0$, and $1.1 \msol$ are labeled with triangles, squares, stars, circles, and diamonds, respectively.  Four metallicities for each set of WD masses and C/O fractions are shown: $0$ (\emph{green}), $0.5$ (\emph{red}), $1$ (\emph{yellow}), and $2 \, Z_\odot$ (\emph{blue}).  The default value of the $^{12}$C$+^{16}$O reaction rate is used in the top panel; the rate is reduced by a factor of $10$ in the bottom panel.  Observational values compiled by \cite{mart17a} are shown in gray; for the remnant G337.2, there is no reliable constraint on the Cr/Fe mass ratio.}
  \label{fig:CrFeCaS}
\end{figure}

We now turn to an exploration of Cr/Fe vs.\ Ca/S mass ratios during the SN remnant phase, motivated by the work of \cite{mart17a}.  While none of these isotopes directly traces neutron-rich nucleosynthesis (Cr and Fe are primarily produced as $^{52}$Fe and $^{56}$Ni, respectively, during the explosion, which have equal numbers of protons and neutrons), the Ca/S ratio does have an inverse correlation with the neutron excess at the time of explosion \citep{de14a,mart17a}.

In Figure \ref{fig:CrFeCaS}, we show our post-processed results for the Cr/Fe mass ratio vs.\ the Ca/S mass ratio during the SN remnant phase, accounting for intermediate decays as before.  All 80 nucleosynthetic calculations are shown, corresponding to five WD masses, four metallicities, initial C/O mass fractions of $50/50$  and $30/70$,  and two choices for the $^{12}$C$+^{16}$O reaction rate: the default \texttt{REACLIB} reaction rate \citep{cf88} and the  rate scaled by a  multiplicative factor, $f_{\rm ^{12}C+^{16}O}=0.1$, as motivated by \cite{mart17a}.  Observational values for five Galactic and LMC remnants from \cite{mart17a} are shown in gray.  The remnant G337.2 does not have constraints on its Cr/Fe mass ratio, so it is shown as a vertical band.

It is clear that only very low metallicity $30/70$ C/O explosions in the top panel are consistent with the observed SN remnants.  As previously mentioned, there is some uncertainty in the fact that some of these remnants may not be old enough to have their entire ejecta traversed by the reverse shock, so that the mass ratios inferred from observations may not be representative of the entire ejecta.  However, the primary discrepancy lies in the Ca/S ratio, and since these IMEs are located in the outer parts of the ejecta, they have likely already been excited by the reverse shock.

Much better agreement is found in the bottom panel, for which the $^{12}$C$+^{16}$O reaction rate is reduced by a factor of 10.  Here, solar and  sub-solar metallicities and C/O ratios of both $50/50$ and $30/70$ match values for observed SN remnants.  Our results are consistent with \cite{mart17a}'s findings; as they explain, a  slower $^{12}$C$+^{16}$O  reaction rate increases the abundance of $^4$He nuclei, which favors the production of isotopes higher in the $\alpha$-chain, and thus a higher Ca/S ratio.

However, the $^{12}$C$+^{16}$O is not actually uncertain to a factor of 10.  Unlike for the typical relatively low-energy stellar case, reaction rates at energies relevant to stellar detonations can be probed in the laboratory.  The burning temperature $\sim \unit[4\E{9}]{K}$ of the carbon detonation yields a Gamow peak of $\unit[7.7]{MeV}$ with width $\unit[3.8]{MeV}$, an energy range at which the cross-section of the $^{12}$C$+^{16}$O reaction has been directly measured.  The \emph{S} factor at the Gamow peak has an experimental uncertainty of only $\sim 50\%$, and its median is actually $\sim 20\%$ higher than the \cite{cf88} value used in \texttt{REACLIB} \citep{patt71a,cuje76a,chri77a,jian07a}. The uncertainty at lower energies within the peak is higher, a factor of $\sim 2$, but since the rate is dominated by the cross-section near the peak's maximum, the rate is only uncertain by $\sim 50\%$ at our temperatures of interest.

Thus, while we do find good agreement with the observed Ca/S ratio in the Tycho and Kepler SN remnants for near-solar metallicities and a reduced $^{12}$C$+^{16}$O  reaction rate, this is not a likely explanation.  Using the default \texttt{REACLIB} \cite{cf88} rate, our sub-$M_{\rm Ch}$ models imply low metallicity progenitors for these remnants.  However, we note that the $M_{\rm Ch}$ models in \cite{mart17a} yield a similar conclusion when the default $^{12}$C$+^{16}$O  reaction rate is used.


\section{Radiative transfer calculations}
\label{sec:rad}

A stringent test of the validity of our sub-$M_{\rm Ch}$ WD detonation models is a comparison to the rich SN Ia observational data sets collected in the past few decades.  To this end, we employ the Monte Carlo radiative transfer code SEDONA \citep{ktn06} to produce synthetic light curves and spectra, which we discuss and compare to observations in the following sections.  These calculations assume the level populations are in local thermodynamic equilibrium (LTE) and that lines are purely absorbing.  Note that we only consider comparisons to ``normal'' SNe Ia, ranging from SN 1991bg-likes to SN 1991T-likes, and not to the peculiar classes of Ca-rich transients (e.g., \citealt{kasl12}) and SNe Iax (e.g., \citealt{fole13a}).


\subsection{Light curves}
\label{sec:lcs}

\begin{figure*}
  \centering
  \includegraphics[width=\textwidth]{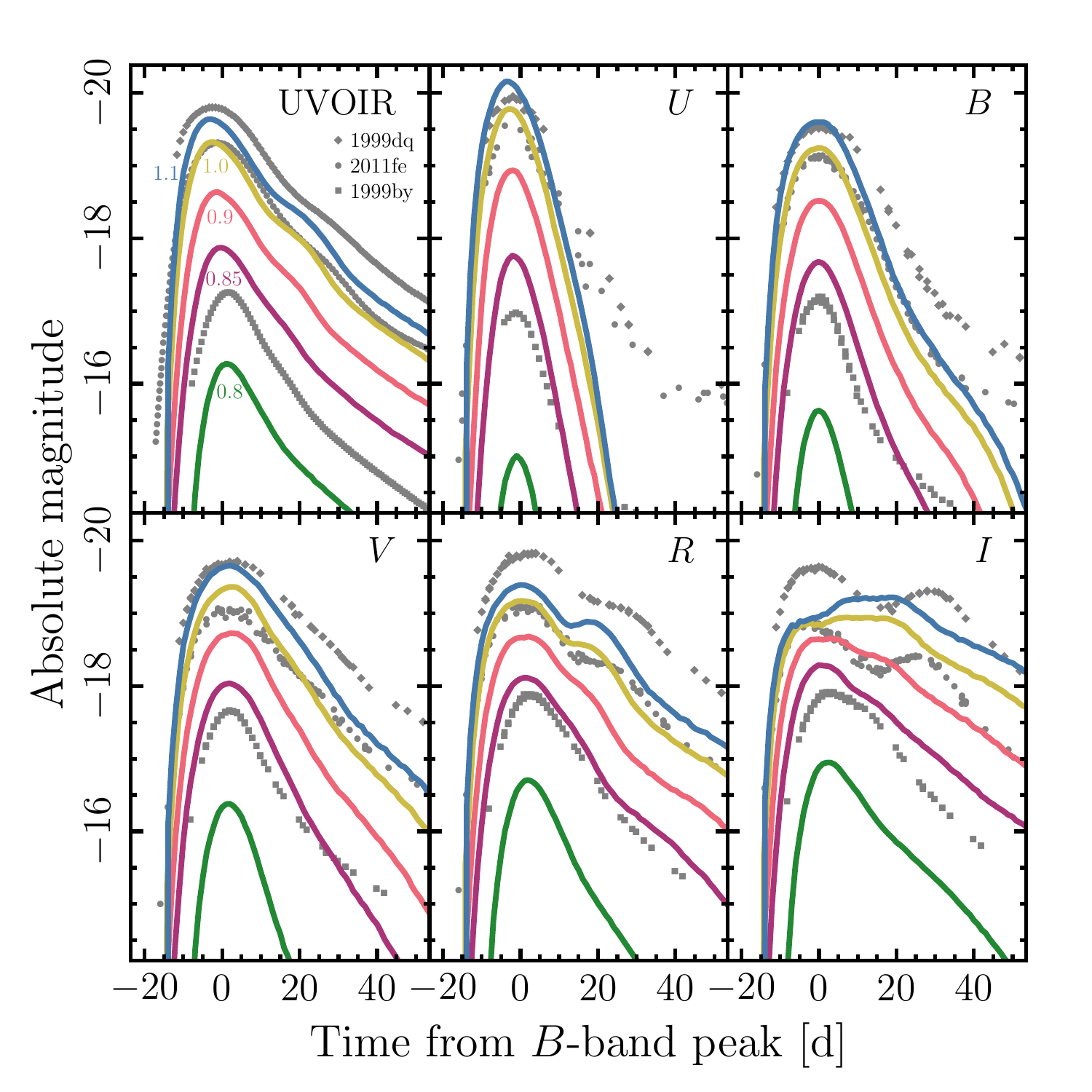}
  \caption{Bolometric and \emph{UBVRI} light curves for five WD masses of $0.8$ (\emph{green}), $0.85$ (\emph{magenta}), $0.9$ (\emph{red}), $1.0$ (\emph{yellow}), and $1.1 \msol$ (\emph{blue}).  The models have an initial C/O mass fraction of $50/50$ and solar metallicity.  Shown for comparison are three well-observed SNe Ia: SN 1999by (\emph{squares}), SN 2011fe (\emph{circles}), and SN 1999dq (\emph{diamonds}).}
  \label{fig:lcs}
\end{figure*}

\begin{figure*}
  \centering
  \includegraphics[width=\textwidth]{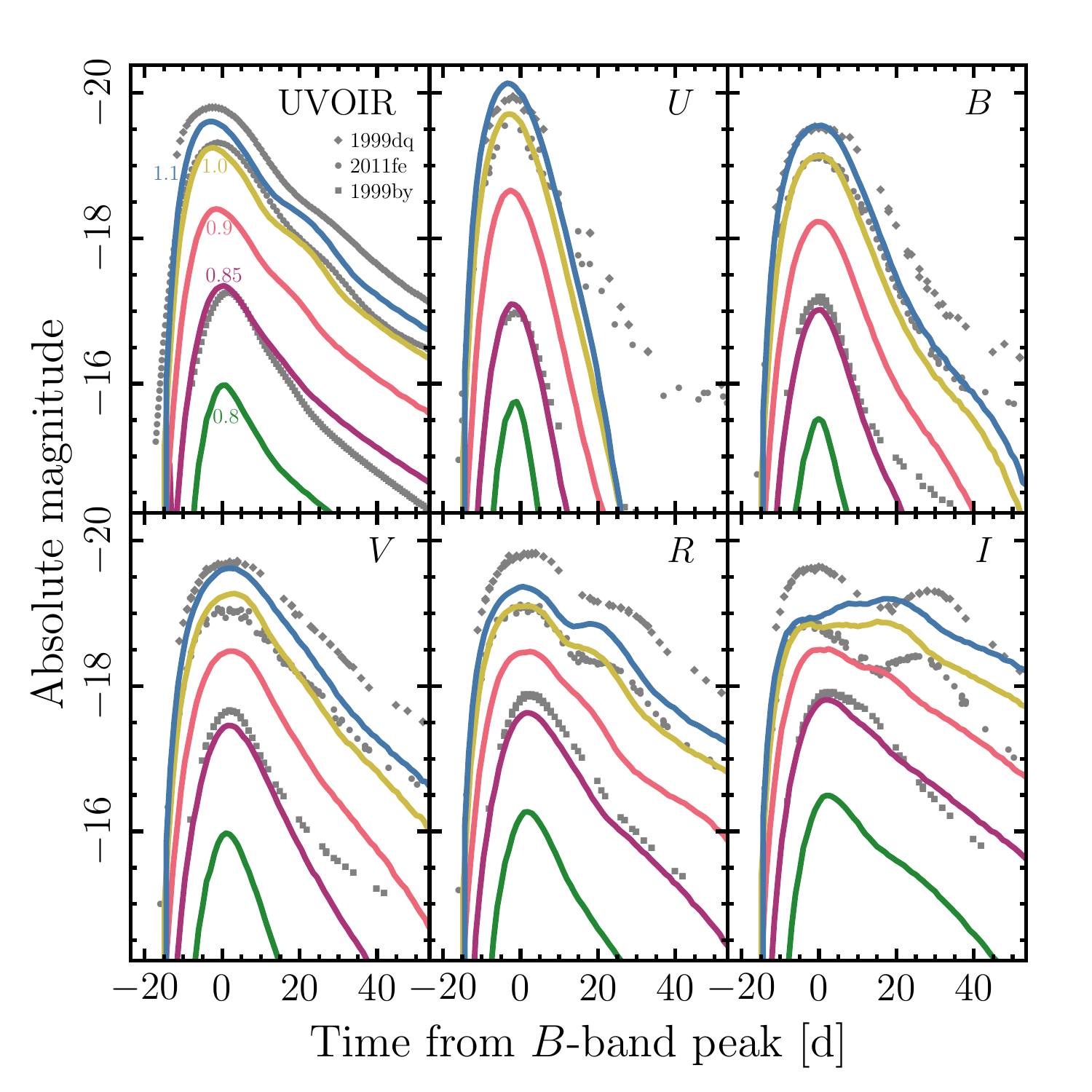}
  \caption{Same as Figure \ref{fig:lcs}, but for an initial C/O mass fraction of $30/70$.}
  \label{fig:lcs_3070}
\end{figure*}

Figures \ref{fig:lcs} and  \ref{fig:lcs_3070} show bolometric and broad-band light curves for post-processed models with solar metallicity and  initial C/O mass fractions of $50/50$ and $30/70$, respectively.  Vega magnitudes are used here and in the following.  Also overlaid for comparison in gray are three well-observed SNe Ia: the subluminous 1991bg-like SN 1999by \citep{garn04,stri05a,gane10a}, the normal SN 2011fe \citep{muna13b,pere13a,tsve13a}, and the over-luminous 1991T-like SN 1999dq \citep{stri05a,jha06b,gane10a}.\footnote{Much of the data used in this work was obtained through \texttt{https://sne.space} \citep{guil17a}.}

The general shapes of our synthetic bolometric light curves show good agreement with observed SNe Ia.  The subluminous SN 1999by is reasonably well fit by our $0.85 \msol$ $30/70$ C/O model, the normal SN 2011fe agrees with the $1.0 \msol$ models, and the over-luminous SN 1999dq is somewhat brighter than our $1.1 \msol$ models.  However, there are some discrepancies in the filtered light curves.  In particular, our synthetic light curves generally fall too rapidly in the $U$ and $B$ bands and remain too bright in the $R$ and $I$ bands.

Our results are in broad agreement with those of \cite{sim10}, although our different nucleosynthetic output precludes an exact comparison.  One notable difference is obvious after $ \unit[30]{d}$, when our bluer light curves deviate from observations, whereas \cite{sim10}'s flatten and provide a better match to observations.  Since this difference persists for higher WD masses, where \cite{sim10}'s and our nucleosynthetic results do not differ drastically, the discrepancy may be due to different treatments of radiative transfer.  As noted by \cite{krom09a} and \cite{sim10}, different radiative transfer codes produce somewhat different light curves for the same input; further work is necessary to ascertain the cause of the mismatch.

\begin{figure}
  \centering
  \includegraphics[width=\columnwidth]{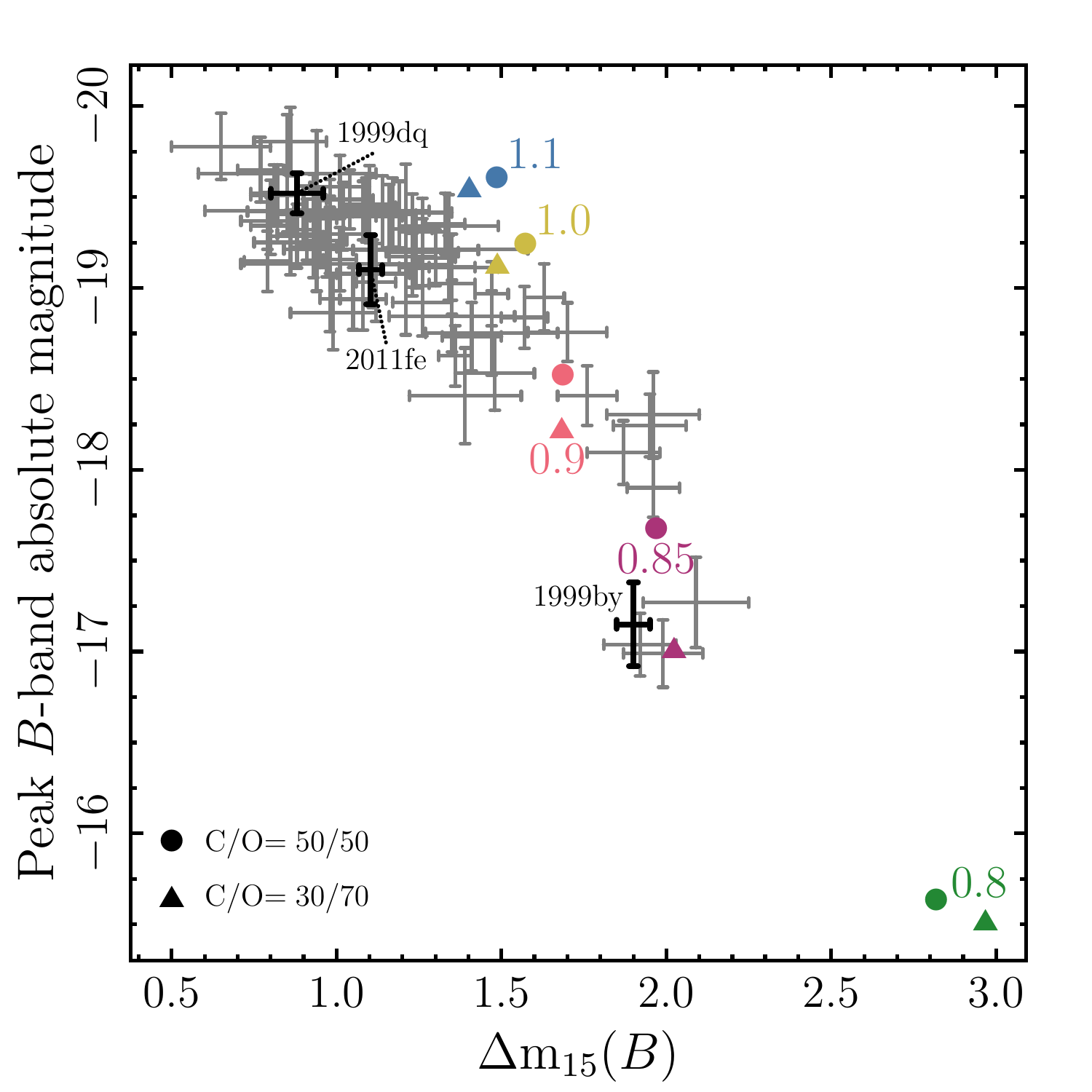}
  \caption{Peak $B$-band absolute magnitude vs.\ $\Delta {\rm m_{15}}(B)$.  Green, magenta, red, yellow, and blue triangles and circles are results from solar metallicity post-processed models, as labeled.  Gray symbols are values taken from the CfA light curve data set \citep{hick09a}, and black error bars are values for SN 1999by, SN 2011fe, and SN 1999dq.}
  \label{fig:phillips}
\end{figure}

The discrepancies in filtered light curves between our results and observations can also be seen in Figure \ref{fig:phillips}, which compares the peak $B$-band absolute magnitude to the decline in magnitudes $\unit[15]{d}$ after maximum, $\Delta {\rm m_{15}}(B)$ \citep{phil93a}.  Our solar metallicity models are shown for our five WD masses and two initial C/O fractions.  Gray error bars are values from the CfA light curve data set \citep{hick09a}, and black symbols are the well-observed SNe Ia used in our light curve comparisons.

Very promisingly, our models reproduce the basic trend of the \cite{phil93a} relation, with more massive WDs yielding brighter SNe Ia that decline more slowly than SNe Ia from less massive WDs.  The agreement is far from exact, though.  Similarly to \cite{sim10}, our high-mass WDs $\ge 1.0 \msol $ lie to the right of the observed relation: they evolve too rapidly compared to observed SNe.

However, as compared to \cite{sim10}, our low-mass WD detonations are brighter and evolve slightly more slowly because of the increased amount of $^{56}$Ni and other IGEs; e.g., our $0.9 \msol$ models are $1.5$ magnitudes brighter and decline $0.1$ magnitudes less after $\unit[15]{d}$ than a $0.9 \msol$ explosion interpolated between their $0.88$ and $0.97 \msol$ models.  Thus, unlike for \cite{sim10}, our $0.85$ and $0.9 \msol$ WD models follow the faint-end slope of the Phillips relation; our $0.85 \msol$ $30/70$ C/O model has a similar peak $B$-band magnitude and $\Delta {\rm m_{15}}(B)$ to those of SN 1991bg-like SNe.

The $B$-band decline rate of our model light curves is highly sensitive to line blanketing effects \citep{kase07a}. The fact that our $1.0$ and $1.1 \msol$ models predict too rapid a decline could be related to limitations in the transport calculations.  In particular, the LTE assumption adopted here, which only approximates the more complex redistribution of photons to longer wavelengths due to fluorescence, may overestimate  the  rate of light curve reddening. As mentioned above, \cite{sim10}'s $U$- and $B$-band  light curves show some late-time flattening, which ours do not.  This difference may be related to their use of a method intended to mimic non-LTE effects.

The importance of these effects is supported by \cite{blon17a}'s non-LTE radiative transfer calculations.  The light curves of their $ \geq 1.0 \msol$ models decline more slowly than ours and those of \cite{sim10}, and they are able to much more closely match the bright end of the Phillips relation.  As discussed in \S \ref{sec:bulkcomp}, the $^{56}$Ni yields do not differ significantly among the various studies at these relatively high masses, and thus the differences in the light curves of the high mass explosions may be attributed to their inclusion of  non-LTE effects.

At the low-mass end, \cite{blon17a}'s non-LTE radiative transfer calculations do not reproduce the Phillips relation, instead yielding light curves that are too dim.  However, our $^{56}$Ni yields are several times higher than theirs in this regime.  It is thus possible that a combination of the nucleosynthesis from our large network, broadened detonation simulations and non-LTE radiation transport calculations will reproduce the entirety of the Phillips relation; such a study is currently underway.

We note that our $0.8 \msol$ models do not appear to match any observed SNe Ia.  As argued by \cite{shen14a}, this may be due to a physical minimum WD mass and associated central density that can be ignited via a converging shock: WDs that are too low in mass cannot explode as double detonation SNe Ia.  Given the qualitative agreement between our $0.85 \msol$ models and SN 1991bg-like SNe, the minimum detonatable WD mass may be $ \simeq 0.85 \msol$.


\subsection{Spectra}
\label{sec:spectra}

\begin{figure}
  \centering
  \includegraphics[width=\columnwidth]{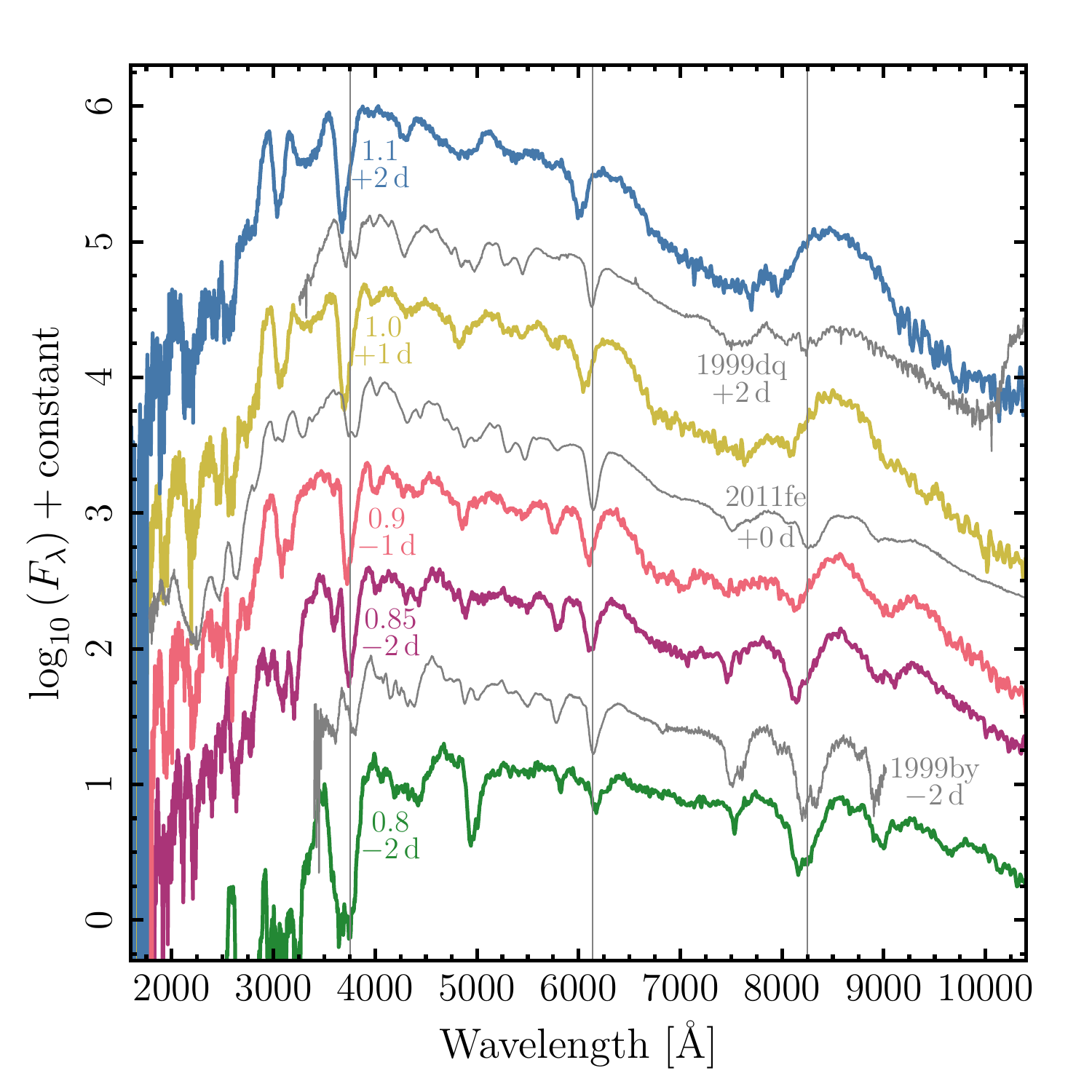}
  \caption{Synthetic and observed spectra near maximum $B$-band magnitude, offset by arbitrary constants.  Green, magenta, red, yellow, and blue lines represent solar metallicity, $50/50$ C/O WDs with masses of $0.8$, $0.85$, $0.9$, $1.0$, and $1.1 \msol$ at $-2$, $-2$, $-1$, $+1$, and $\unit[+2]{d}$ from $B$-band maximum, respectively.  Observed spectra for SN~1999by at $\unit[-2]{d}$, SN~2011fe at $\unit[+0]{d}$, and SN~1999dq at $\unit[+2]{d}$ are shown in gray.  Vertical lines are located in the Si {\sc ii} $ \lambda$6355 and Ca {\sc ii} H\&K and near-IR triplet regions to help guide the eye.}
  \label{fig:spectra}
\end{figure}

\begin{figure}
  \centering
  \includegraphics[width=\columnwidth]{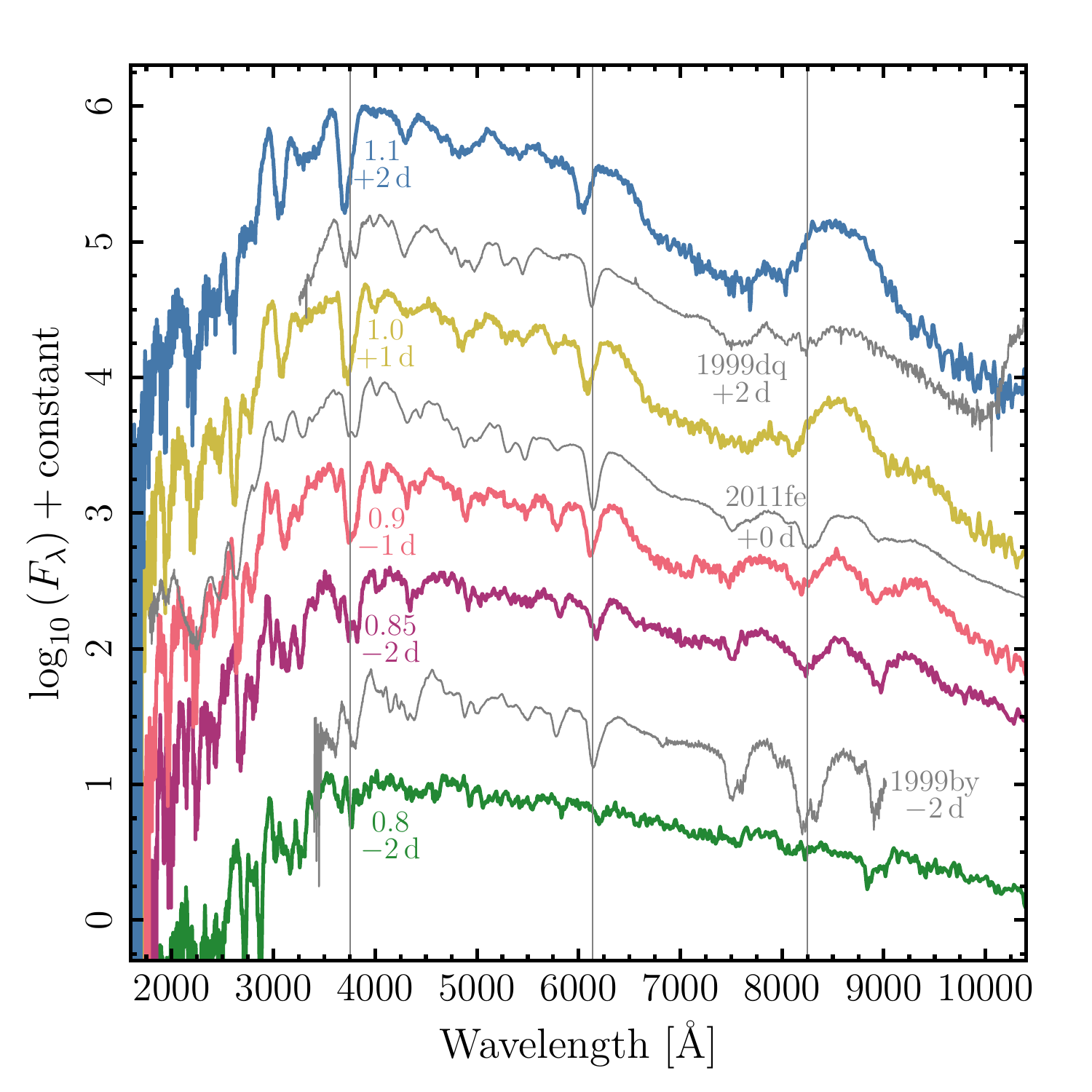}
  \caption{Same as Figure \ref{fig:spectra}, but for an initial C/O mass fraction of $30/70$.}
  \label{fig:spectra_3070}
\end{figure}

Figures \ref{fig:spectra} and \ref{fig:spectra_3070} compare synthetic near-maximum spectra of our five solar metallicity WD models with initial C/O mass fractions of $50/50$ and $30/70$, respectively, to spectra of SN~1999by, SN~2011fe, and SN1999dq at $-2$, $+0$, and $\unit[+2]{d}$ from $B$-band maximum.   While detailed features are not matched precisely, the overall agreement is promising.  Our synthetic spectra show the hallmark attributes of SNe Ia -- strong Si, S, Ca, and Fe features  -- with reasonable correspondence to observed line strengths.  The $\unit[4000-4500]{\AA}$ Ti \textsc{ii} trough characteristic of subluminous SNe Ia is also partially reproduced in our least massive $0.8$ and $0.85 \msol$ models.

One of the most significant discrepancies between our synthetic spectra and observations are the IME velocities, particularly for the more massive WDs $\ge 1.0 \msol$ and brighter observed SNe.  At $B$-band maximum, the Si and Ca velocities of our $1.0$ and $1.1 \msol$ explosions are several thousand $\unit[]{km \, s^{-1}}$ higher than observed.  It is possible that non-LTE calculations will alleviate this discrepancy, as \cite{blon17a}'s sub-$M_{\rm Ch}$ spectra possess appropriate line velocities.

A resolution to this issue may also lie in future multi-dimensional explosion studies of the double detonation scenario.  In the converging shock variant of the double detonation, a helium shell detonation propagates around the WD's surface and launches an oblique shock into the core that focuses its energy near the center and ignites the carbon detonation \citep{livn90,fhr07,fink10,shen14a}.  This inwardly propagating shock may tamp the outgoing core detonation somewhat and reduce the velocities of the outermost ejecta where the IME features form.

In the edge-lit double detonation variation, the helium shell detonation transitions into a carbon-powered detonation as soon as it encounters the WD core \citep{taam80a,taam80b,nomo82b,wtw86}.  Thus, for one hemisphere of the WD, the carbon detonation actually moves inwards initially, so that when pressure forces cause the ejecta to rebound outwards, the outermost ejecta velocity will be similarly limited.  For the opposite hemisphere, tamping of the outer ejecta may still occur if the helium shell detonation races ahead and reaches the opposite pole before the carbon detonation traverses the WD core.

\cite{krom10} performed multi-dimensional converging shock double detonation simulations that are similar to our planned future calculations.  They found that standard SN Ia light curves and spectra are only produced if the helium shells are heavily polluted by $^{12}$C ($\sim 30\%$ by mass).  However, the minimum detonatable helium shell masses found by \cite{shen14b} are an order of magnitude smaller than those used by \cite{krom10}.  We remain hopeful that these much smaller realistic helium shells will still lead to tamping of the bulk ejecta's velocities without adversely affecting the light curves and overall spectra.  There is also the intriguing possibility that these minimal helium shells, which only produce Si and Ca ashes \citep{moor13a,shen14b}, will also explain the high-velocity ($\gtrsim \unit[2\E{4}]{km \, s^{-1}}$) features seen in most SNe Ia \citep{chil14a,magu14a,silv15a}.


\section{Conclusions}
\label{sec:conc}

Motivated by discrepancies in the literature and a need for detailed nucleosynthetic data, we have revisited simulations of bare sub-$M_{\rm Ch}$ C/O WD detonations.  We use a detonation-broadening scheme in a $41$-isotope hydrodynamical simulation to spatially resolve the detonation structure and show convergence of the results with increasing resolution.  These results are then post-processed with a $205$-isotope nuclear reaction network.  Our bulk nucleosynthetic results confirm recent work by \cite{moll14a} and disagree with the studies by \cite{sim10} and \cite{blon17a}, especially for low-mass WDs.  Our examination of neutron-rich nucleosynthesis counters some of the previous claims for $M_{\rm Ch}$ explosions from the solar abundance of Mn and from observations of SN remnants, but future work is necessary to resolve remaining tensions.

The synthetic light curves and spectra of our simulations show promising similarities to observations.  We find that typical SN 2011fe-like SNe Ia can be produced by the detonations of $1.0 \msol$ WDs, which are more numerous than the $1.1 \msol$ WDs required by \cite{sim10} to produce typical SNe Ia.  This lower mass requirement will increase binary population synthesis rates of SNe Ia from the dynamically driven double degenerate double detonation scenario as well as from violent double WD mergers that directly ignite carbon \citep{pakm10,kash15a,tani15a,sato15a}.  This revision of the mapping of detonating WD mass to SN Ia luminosity is a necessary input to recent work on the evolution of the SN Ia luminosity function \citep{shen17c}, another piece of evidence that the bulk of SNe Ia arise from sub-$M_{\rm Ch}$ WD explosions.

The peak luminosities and evolutionary timescales of our radiative transfer results are correlated in a similar way to the observed \cite{phil93a} relation, and the spectral features and line ratios are in general agreement with observed spectra.  However, there is significant disagreement in the line centers of the IMEs and in the evolutionary timescales for the high-mass WD explosions.  We are hopeful that future calculations building on this work, including more precise treatments of radiation transport, will resolve these discrepancies.  We will also verify the yields produced by this front-broadening scheme using comparisons to fully-resolved calculations of the microscopic structure of steady-state detonations.

Future work will also include multi-dimensional simulations with the very low-mass detonatable helium shells found by \cite{shen14b}.  We will employ a similar detonation-broadening scheme, which provides an artificial but numerically resolved model that we expect to give resolution-independent results for modest grid resolutions.  The inward shock from the helium detonation has the potential to tamp the IME velocities and bring our radiative transfer results into agreement with observations, and the ashes from the helium-burning may also provide a satisfying explanation for the high-velocity features observed in most SNe Ia.

While much future work remains to be done, this study has bolstered the potential for sub-$M_{\rm Ch}$ WD detonations in double WD binaries to explain most SNe Ia.   Theoretical and observational studies are beginning to converge, and we are hopeful that the solution to the SN Ia progenitor mystery now lies within reach.


\acknowledgments

We thank Tony Piro for his encouragement, Stefan Taubenberger for providing access to data, and Carles Badenes, St\'{e}phane Blondin, Hector Mart\'{i}nez-Rodr\'{i}guez, Alison Miller, R\"{u}diger Pakmor, Stuart Sim, and Frank Timmes for helpful discussions.  KJS, DMT, and BJM received support from the NASA Astrophysics Theory Program (NNX15AB16G and NNX17AG28G). DK is supported in part by a DOE Office of Nuclear
Physics Early Career Award, and by the Director, Office of Energy
Research, Office of High Energy and Nuclear Physics, Divisions of
Nuclear Physics, of the US DOE under Contract No.\
DE-AC02-05CH11231.  This research used the Savio computational cluster resource provided by the Berkeley Research Computing program at the University of California, Berkeley (supported by the UC Berkeley Chancellor, Vice Chancellor of Research, and Office of the CIO), high-performance computing resources provided by the University of Alabama, and resources of the National Energy Research Scientific Computing Center, a DOE Office of Science User Facility supported by the Office of Science of the US DOE under Contract No.\ DE-AC02-05CH11231.

\software{MESA (v8845; \citealt{paxt11,paxt13,paxt15a}),
FLASH (v4.2.2; \citealt{fryx00,dube09a}), SEDONA \citep{ktn06}}



\end{document}